\documentclass[aps,prx,longbibliography,amsmath,amssymb,floatfix,twocolumn,groupedaddress]{revtex4-1}
\pdfoutput=1
\usepackage{graphicx}
\usepackage{hyperref}
\usepackage{dsfont}
\usepackage[lofdepth,lotdepth,caption=false]{subfig}
\renewcommand{\vec}[1]{\mathbf{#1}}

\begin{document}

\title{Onsager rule,  quantum oscillation frequencies, and the density of states in the mixed-vortex state of cuprates}

\author{Zhiqiang Wang}
\affiliation{Department of Physics and Astronomy, University of
California Los Angeles, Los Angeles, California 90095-1547}
\author{Sudip Chakravarty}
\affiliation{Department of Physics and Astronomy, University of
California Los Angeles, Los Angeles, California 90095-1547}

\date{\today}

\begin{abstract}
The Onsager rule determines the frequencies of quantum oscillations in  magnetic fields.   We show that this rule remains  intact to an excellent approximation in the mixed-vortex state of the underdoped cuprates even though the Landau level index $n$ may be fairly low, $n\sim 10$. The models we consider are fairly general, consisting of a variety of density wave states combined with $d$-wave superconductivity within a mean field theory. Vortices are introduced as quenched disorder and averaged over many realizations, which can be considered as snapshots of a vortex liquid state.  We also show that  the oscillations ride on top of a  field independent density of states, $\rho(B)$,  for higher fields. This feature appears to be consistent with recent specific heat measurements [C. Marcenat, et al. Nature Comm. {\bf 6}, 7927 (2015)]. At lower fields we  model the system as an ordered vortex lattice, and show that its density of states follows a dependence  $\rho(B)\propto \sqrt{B}$ in agreement with the semiclassical results [G. E. Volovik, JETP Lett. {\bf 58}, 469 (1993)].
\end{abstract}


\maketitle
\section{Introduction}
 A  breakthrough in the area of cuprate superconductivity  is the  observation
of quantum oscillations in  cuprates~\cite{Doiron2007,Sebastian2008}. In these experiments a strong magnetic field
is applied to suppress the superconductivity, which  most likely reveals the  ground state~\cite{Chakravarty2008b} without superconductivity. However, the understanding of this ``normal state'' may be a crucial ingredient in the theory high temperature superconductivity.
Standing in the way are at least two important issues: (1) Does the quantum oscillation frequencies substantially  deviate from the classic Onsager rule  for which the oscillation frequency $F=(\hbar c/2\pi e) A(\epsilon_{F})$, where  $A(\epsilon_{F})$  is equal to the extremal Fermi surface area normal to the magnetic field? If so, it would lead to considerable uncertainty in the interpretation of the experiments. (2) Do the oscillations ride on top of a magnetic field dependence of the density of states(DOS) $\rho(B)\sim \sqrt{B}$?~\cite{Riggs2011} If so, it might indicate the presence of superconducting fluctuations even in high magnetic fields at zero temperature, $T=0$, from an extrapolation of a result of Volovik,~\cite{Volovik1993} which is supposed to be asymptotically true as $B\to 0$. Therefore  high field behavior requires careful analyses.
Quantum oscillations require the existence of Landau levels. If this is true, they might indicate
the existence of normal Fermi liquid quasiparticles~\cite{Chakravarty2011}.

It has been argued from a theoretical analysis that the Onsager rule could be violated by as much as $30\%$~\cite{Chen2009}. We find that under reasonable set of parameters, to be defined below, the violation is miniscule, $\sim 10^{-4}$. Even for extreme situations discussed in Ref.~\citenum{Chen2009}, it is less than $2\%$.
If we are correct, one can use the Onsager rule to interpret the experiments with impunity. The second encouraging result is that $\rho(B)$ saturates in the regime where oscillations are present. We interpret this to mean that there are generically no superconducting fluctuations in  high fields. A recent specific heat measurement~\cite{Marcenat2015} shows that the specific heat indeed saturates at high fields, signifying that the normal state is achieved.

To put our paper in the context, note that  in conventional $s$-wave superconductors, previous work has  shown that for  higher  Landau level indices, and within coherent potential approximation,
vortices mainly damp the oscillation amplitude, but the shift in the oscillation frequency~\cite{Stephen1992,Maki1991} is negligible; however, for $d$-wave underdoped cuprate superconductors with small coherence length and high fields with Landau level indices $~\sim 10$ this
calculation should not hold~\cite{Chen2009}.  A more recent semi-classical analysis based on an ansatz of  {\em gaussian phase fluctuations} of the $d-$wave pairing ~\cite{Banerjee2013} indicates that the oscillation frequency is unchanged, as here. However, relatively undamped quantum oscillations  riding on top of $\sqrt{H}$ was found in this dynamic Gaussian ansatz that does not account for vortices, which must necessarily be present, as in Ref.~\cite{Chen2009}, and the branch cuts introduced by the vortices must also be taken into account.

We consider the  vortices explicitly in the Bogoliubov-de Gennes (BdG) Hamiltonian, as  in Ref.~\cite{Chen2009}, and model the vortex liquid state as quenched, randomly distributed vortices, paying special attention to branch cuts. There are other important differences as well, as we shall discuss below.
To have a complete picture, we also consider the low field regime where the quantum oscillations disappear. In this regime the vortices arrange themselves into a vortex solid state and should be modeled as an ordered lattice instead. We compute the DOS of such vortex lattices explicitly and find that  $\rho(B)\propto \sqrt{B}$ in the asymptotically low field limit, consistent with Volovik's semiclassical analysis~\cite{Volovik1993}.

In Section II we define the model Hamiltonian that includes $d$-wave superconducting order parameter as well as a variety of density wave states. Our numerical method, the recursive Green function method adapted for the present problem is discussed in Sec. III. The results are discussed in Sec. IV and Sec. V contains discussion. There are three appendices.

\section{ The Model Hamiltonians}
The starting point is  the Bogoliubov-de Gennes (BdG) Hamiltonian
\begin{equation}
      \mathcal{H}=\left(\begin{array}{cc} H-\mu & \Delta_{ij} \\
      \Delta^{\dagger}_{ij} & -H+\mu \end{array}\right )
\end{equation}
defined on a square lattice. Here $\mu$ is the chemical potential. $H$ is the Hamiltonian that describes the normal state electrons; while the off-diagonal pairing term $\Delta_{ij}$ defines the superconducting order parameter.
For simplicity, we ignore self consistency, as we believe that it cannot change the major striking conclusions.

\subsection{The diagonal component $H$}
Besides the hopping parameters, the normal state Hamiltonian $H$  contains a variety of mean field order parameters defined below. Although  many different orders are suggested to explain the normal state of the high-$T_c$ superconductivity, for our purposes it is sufficient to consider three different types: a period-2 $d-$density wave (DDW)~\cite{Chakravarty2008a,Chakravarty2011,Eun2012}, a bi-directional charge density wave (CDW), and a period$-8$ DDW model~\cite{Eun2012}. We believe that our major conclusions in this paper do not depend on the nature of the density wave that is responsible for Fermi surface reconstruction. The DDW is argued to be able to account for many features of  quantum oscillations, as well as the pseudogap state~\cite{Laughlin2014a,Laughlin2014b} in the cuprates. Among the many different versions of density waves of higher angular momentum~\cite{Nayak2000}, the simplest period-2 singlet DDW, also the same as staggered flux state in Ref.~\cite{Chen2009}, and also a period$-8$ DDW order, proposed by us previously to explain quantum oscillations\cite{Eun2012}, have been chosen here for illustration.

Recently a bi-directional CDW has been observed ubiquitously in the underdoped cuprates~\cite{Wu2011,Wu2013,Ghiringhelli2012,Chang2012,Blackburn2013}.  It has ordering wavevectors $\vec{Q}_1\approx \frac{2\pi}{a}(0.31,0)$ and $\vec{Q}_2\approx \frac{2\pi}{a}(0,0.31)$, which are incommensurate. This order has also been used to explain the Fermi surface reconstructions and quantum oscillation experiments~\cite{Sebastian2014,Allais2014}, although a recent numerical work~\cite{Zhang2015} has demonstrated that the strict incommensurability of the CDW can destroy {\em strict} quantum oscillations completely. For  the purpose  of illustration we chose, instead,  commensurate  vectors $\vec{Q}_1= \frac{2\pi}{a}(\frac{1}{3},0)$ and $\vec{Q}_2= \frac{2\pi}{a}(0,\frac{1}{3})$.

Therefore without the magnetic field, $B$,  $H$ is given by
\begin{eqnarray}
H& = -t \sum\limits_{\langle i,j \rangle}   c_{\vec{r}_i}^{\dagger} c_{\vec{r}_j}\,  + t^{\prime} \sum\limits_{\langle \langle i,j \rangle \rangle} c_{\vec{r}_i}^{\dagger} c_{\vec{r}_j} \nonumber \\
 & -t^{\prime\prime} \sum\limits_{\langle \langle \langle i,j \rangle \rangle \rangle} c_{\vec{r}_i}^{\dagger} c_{\vec{r}_j}+ \mathrm{h.c.}+H_{\mathrm{d. w.}},
\end{eqnarray}
where $t$,$t^{\prime}$, and $t^{\prime\prime}$ are the 1st, 2nd and the 3rd nearest neighbor hopping parameters respectively. $H_{\mathrm{d.w.}}$ is various density wave orders specified below.
The external uniform magnetic field $\vec{B}=B\hat{z}$ is included into $H$ via the Peierls  substitution: $ c_{\vec{r}_i}^{\dagger}c_{\vec{r}_j} \Rightarrow \exp[{-i\frac{e}{\hbar c}\int^{\vec{r}_i}_{\vec{r}_j}\vec{A}\cdot d\vec{l}}] c_{\vec{r}_i}^{\dagger}c_{\vec{r}_j} $ with the vector potential $\vec{A}=B\, x \,\hat{y}$ chosen, for simplicity,  in the Landau gauge.

\begin{enumerate}

\item {\bf Two-fold DDW order}
\begin{gather}
H_{\mathrm{d.w.}}= \sum_{\vec{r}_i,\delta} i \frac{W_0}{4} (-1)^{x_i+y_i} \, \eta_{\delta} \, c_{\vec{r}_i+\delta}^{\dagger}c_{\vec{r}_i},
\end{gather}
where $\delta=\hat{x},\hat{y}$ denote the two nearest neighbors. $\eta_{\delta}=1$ for $\delta=\hat{x}$ while $\eta_{\delta}=-1$ for $\delta=\hat{y}$ indicates the DDW order has a local $d-$wave symmetry.

\item
{\bf Bi-directional CDW order}
\begin{align}
H_{\mathrm{d.w.}} & = V_{\mathrm{c}} \sum_{\vec{r}_i,\delta} \eta_{\delta} \; \{ \cos[\vec{Q}_1\cdot (\vec{r}_i +\delta/2)] \nonumber \\
& +\cos[\vec{Q}_2\cdot (\vec{r}_i+\delta/2)]\} \;  c_{\vec{r}_i+\delta}^{\dagger}c_{\vec{r}_i}.
\end{align}
Again $\eta_{\delta}=\pm 1$ is the local $d-$wave symmetry factor of the CDW order.
\item
{\bf Period-8 DDW order}
\begin{align}
H_{\mathrm{d.w.}} & =\sum_{\vec{k}}  i G_k \,c_{\vec{k}}^{\dagger} c_{\vec{k}+\vec{Q}} + V_c\,c_{\vec{k}}^{\dagger} c_{\vec{k}+2\vec{Q}}+\mathrm{h.c.} \;. \label{eq:p8hamiltonian}
\end{align}
where the $ i G_k$ term is the period$-8$ DDW order
\begin{gather}
<c^{\dagger}_{\vec{k}^{\prime}}c_{\vec{k}}> = i G_{\vec{k}} \, \delta_{\vec{k}^{\prime},\vec{k}+\vec{Q}} - i G_{\vec{k}^{\prime}}  \,\delta_{\vec{k},\vec{k}^{\prime}+\vec{Q}}
\end{gather}
Here $G_{\vec{k}}=(W_{\vec{k}} - W_{\vec{k}+\vec{Q}})/2$ with the DDW gap $W_{\vec{k}} = \frac{W_0}{2}(\cos k_x - \cos k_y)$, and the ordering wavevector is $\vec{Q}=(\frac{3\pi}{4a},\frac{\pi}{a})$. The $V_c$ term in Eq.~\eqref{eq:p8hamiltonian} represents a period$-4$ unidirectional CDW with an ordering wavevector $2\vec{Q}$, which is consistent with the symmetry of the period$-8$ DDW order. Notice that this CDW is different from the bi-directional CDW we considered in the previous section. Experimentally whether the observed CDW in cuprates is unidirectional or bi-directional is still not fully resolved.

Fourier transformed to the real space, the Hamiltonian $H_{\mathrm{d.w.}}$ becomes
\begin{eqnarray}
\;\;\;\;\;\:&&H_{\mathrm{d.w.}}=\sum_{\vec{r},\vec{r}^{\prime}}  i\, \frac{W_0}{2} \;  \sin \frac{\vec{Q}\cdot(\vec{r}-\vec{r}^{\prime})}{2} \sin \frac{\vec{Q}\cdot(\vec{r}+ \vec{r}^{\prime})}{2} \nonumber \\ && \quad \times \{\delta_{\vec{r},\vec{r}^{\prime}+a\hat{x}} + \delta_{\vec{r},\vec{r}^{\prime}-a\hat{x}}-\delta_{\vec{r},\vec{r}^{\prime}+a\hat{y}}-\delta_{\vec{r},\vec{r}^{\prime}-a\hat{y}}\}   c^{\dagger}_{\vec{r}}c_{\vec{r}^{\prime}} \nonumber \\
  && \quad +2\,V_{c} \sum_{\vec{r}} \cos[2\, \vec{Q}\cdot \vec{r}] \; c^{\dagger}_{\vec{r}}c_{\vec{r}}.  \label{eq:pehamiltonian-2}
\end{eqnarray}
In the above, $W_0$ controls the overall magnitude of the period-8 DDW order parameter, $i \; \sin \frac{\vec{Q}\cdot(\vec{r}-\vec{r}^{\prime})}{2}$ indicates that the order is a current, $\sin \frac{\vec{Q}\cdot(\vec{r}+ \vec{r}^{\prime})}{2}$ shows that the magnitude of this order is modulated with a wavevector $\vec{Q}$, and the last factor in the curly bracket $\{ ...\}$ explicitly exhibits its local $d-$wave symmetry. If $\vec{Q}=(\pi/a,\pi/a)$, then the order reduces to the familiar two-fold DDW order. In that case $|\sin \frac{\vec{Q}\cdot(\vec{r}+ \vec{r}^{\prime})}{2}|=1$ and the order parameter magnitude is a constant. The last term in Eq.~\eqref{eq:pehamiltonian-2} gives the $2\,\vec{Q}$ charge modulation. Note that this CDW is defined on sites, differing from the bi-directional CDW defined on bonds.
\end{enumerate}

\subsection{The off-diagonal component $\Delta_{ij}$}
The off-diagonal pairing term $\Delta_{ij}$ in the BdG Hamiltonian is defined on each bond connecting two nearest neighboring sites $i$ and $j$.  $\Delta_{ij}=|\Delta_{ij}|e^{i\theta_{ij}} \; \eta_{ij}$, where $\eta_{ij}=+1$ if the bond
is along $x-$direction and $\eta_{ij}=-1$ if it is along $y-$direction so that $\Delta_{ij}$ has a local $d-$wave symmetry. The pairing amplitude is taken to be
\begin{equation}\label{eq:pairing}
|\Delta_{ij}|= \Delta \frac{r_{\mathrm{eff}}}{\sqrt{r_{\mathrm{eff}}^2+\xi^2}}
\end{equation}
where $\Delta$ is the pairing amplitude far away from any vortex center. $\xi$ is the vortex core size. In our calculation $\xi=5a$ is adopted, where $a$ is the lattice spacing. In the presence of a single vortex, $r_{\mathrm{eff}}$ in the above is simply the distance from the center of our bond $\frac{\vec{r}_{i}+\vec{r}_j}{2}$ to the center of that vortex. While in the presence of multiple vortices, following the ansatz used in Ref.~\cite{Chen2009} we choose $(\frac{\xi}{r_{\mathrm{eff}}})^{q}=\sum_{n} (\frac{\xi}{r_n})^{q}$ where $r_n$ is the distance from the bond center to the $n$th vortex center and $q > 0$ is some real number.

In this ansatz, $r_{\mathrm{eff}}$ is a monotonic increasing function of the parameter $q$ for a given vortex configuration. Therefore if $q$ is large, the calculated $r_{\mathrm{eff}}$ as well as $|\Delta_{ij}|$ is also larger, which means the vortex scattering is stronger. However our conclusions do not depend on the different choices of $q$ (for more details see the appendix section~\ref{sec:ansatz}). Therefore in this  paper, if not specified otherwise, $q=2$ will be chosen.

The bond phase variable $\theta_{ij}$ contains the information of our quenched random vortex configuration, but for the purpose of our calculation we need the site phase variables. We use the ansatz for $\theta_{ij}$ given in Refs.~\cite{Melikyan2006,Vafek2006}
\begin{equation}
e^{i\theta_{ij}} = e^{i \frac{\phi_i+\phi_j}{2}} \, \mathrm{sgn}[\cos \frac{\phi_i-\phi_j}{2}]
\end{equation}
where $\phi_i$ is the pairing order parameter phase field defined on a site. In the above, without the ``sgn[...]" factor $\theta_{ij}$ is simply the arithmetic mean of  $\phi_i$ and $\phi_j$. However using $\theta_{ij}=\frac{\phi_i+\phi_j}{2}$ is not enough because whenever the bond $\overline{ij}$ crosses a vortex branch cut, the phase factor $e^{i\theta_{ij}}$ will be incorrect and different from the correct one by a minus sign. This can be corrected by  the additional ``sgn[...]" factor(see the appendix section~\ref{sec:thetaij}).

\begin{figure}[ht]
  \centering
  \includegraphics[width=0.4\textwidth]{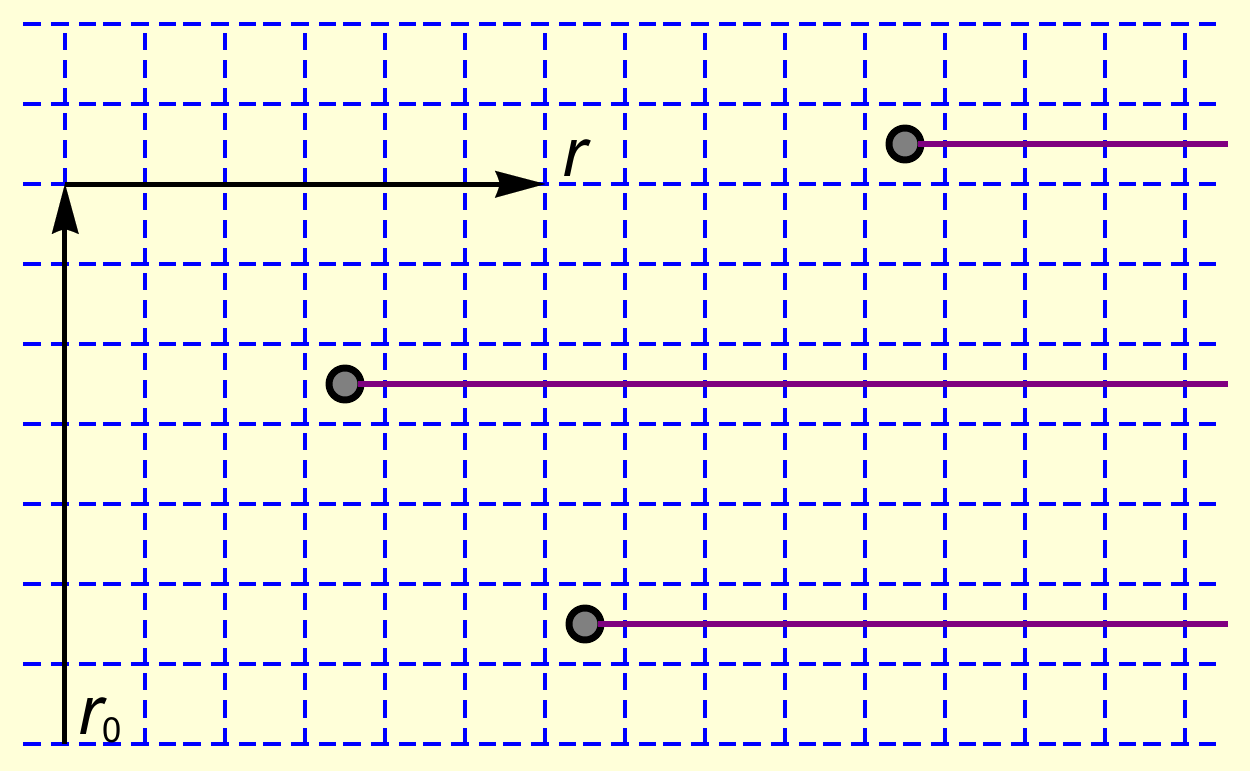}
  \caption{Illustrations of the vortices (circle) on the lattice (dashed lines). The arrows show the path of the integral we have chosen in defining our phase field $\phi(\vec{r})$. To make this phase definite, the branch cuts of all the vortices are chosen to extend from the vortex center to the positive infinity ($x=\infty$), represented by the magenta horizontal lines. }
  \label{fig:vortexbranchcut}
\end{figure}

Then $\phi_i$ can be further computed from the superfluid velocity field $\vec{v}_s(\vec{r}_i)$ by
\begin{eqnarray}
      \phi_i-\phi_0 & = \int_{\vec{r}_0}^{\vec{r}_i} [\frac{m^{*} \vec{v}_s(\vec{r})}{\hbar}+\frac{e^{*}}{\hbar c}\vec{A}(\vec{r})] \cdot d\vec{l}
\end{eqnarray}
with $m^{*}= 2 m$ and $e^{*}=-2 e$ are the mass and the charge of the Cooper pairs respectively. The path for this integral is chosen such as to avoid the branch cuts of all the vortices so that the phase field $\phi_i$ is single valued on every site, as illustrated in Fig.~\ref{fig:vortexbranchcut}.

We still need to compute the superfluid velocity $\vec{v}_s(\vec{r})$. This can be done by following Ref.~\cite{Tinkham1996}
\begin{align}
m \vec{v}_s(\vec{r}) & = - i \pi \hbar \int \frac{d^2 k}{(2\pi)^2} \frac{\vec{k}\times \hat{z}}{k^2+\lambda^{-2}} \sum_n  e^{i\, \vec{k}\cdot (\vec{r}-\vec{R}_n)} \label{eq:vs},
\end{align}
where $m$ is the electron mass, $\lambda$ is the penetration depth, and $\vec{R}_n$ gives the $n$th random vortex position. In this integrand, because $\vec{k}\times\hat{z}$ is odd in $\vec{k}$, only the imaginary part of $e^{i\, \vec{k}\cdot (\vec{r}-\vec{R}_n)}$ will survive after the integration, so the whole expression on the right hand side becomes real. We also make an approximation $\lambda=\infty$ so that we can ignore the $\lambda^{-2}$ term in the denominator. This is equivalent to replacing the magnetic field $\vec{B}(\vec{r})$ by its spatial average, which is equal to the external magnetic field $\vec{B}=B\hat{z}$. It is a good approximation when $B\gg H_{c1}$, where $H_{c1}$ is the lower critical field. This condition is well satisfied in the quantum oscillation experiments of cuprates. Also this approximation is consistent with our initial choice of the vector potential $\vec{A}=Bx\hat{y}$, given completely by the applied external field $\vec{B}$.

For our square lattice calculation we discretize the above $\vec{k}$ integral and choose  $2\pi/\xi$ as its upper cutoff, since the vortex is only well defined over a length scale larger than the vortex core size $\xi$. Therefore in the limit $\lambda \gg \xi > a$, $\vec{v}_s(\vec{r})$ can be rewritten as follows
\begin{gather}
m \, \vec{v}_s(\vec{r}) =  \frac{\pi \hbar}{LM\, a^2} \sideset{}{'}\sum_{(k_x,k_y)} \frac{\vec{k}\times \hat{z}}{k^2} \sum_n \sin[\vec{k}\cdot(\vec{r}-\vec{R}_n)]. \label{eq:vs2}
\end{gather}
In this summation $k_x=-\frac{2\pi}{\xi},-\frac{2\pi}{\xi}+\frac{2\pi}{La}, .... ,\frac{2\pi}{\xi}-\frac{2\pi}{La},\frac{2\pi}{\xi}$, $k_y=-\frac{2\pi}{\xi},-\frac{2\pi}{\xi}+\frac{2\pi}{Ma}, .... ,\frac{2\pi}{\xi}-\frac{2\pi}{Ma},\frac{2\pi}{\xi}$. The prime superscript in the summation means the point $(k_x,k_y)=(0,0)$ is excluded to be consistent with our approximation $\lambda=\infty$.

\section{The recursive Green function}
Given the  BdG Hamiltonian $\mathcal{H}$ defined above, we use the recursive Green's function method~\cite{Stauffer1996} to compute the local DOS(LDOS). We attach our central system, which has a lattice size $L\times M$, to two semi-infinite leads in the $\pm x$ directions. The leads are normal metals described by $t,t^{\prime},t^{\prime\prime}$ only. Then we can compute the retarded Green's function $G_i(j,j^{\prime}; E+i\delta)$ at an energy $E$ for the ${i}$th principal layer(see the appendix section~\ref{sec:rGF}). Here each ${i}$th principal layer contains two adjacent columns of the original square lattice sites. So there are $L/2$ principal layers and each of them contains $2M$ number of sites. Therefore $G_i(j,j^{\prime}; E+i\delta)$ is a $4M\times 4 M$ matrix, with $j,j^{\prime}=1,2,...,4M$, because it has both an electron part and a hole part. In calculating the LDOS at the ${j}$th site of the ${i}$th layer only the imaginary part of the $j$th diagonal element in the electron part of $G_i$ is included. This is equivalent to treating the random vortices as some off-diagonal scattering centers for the normal state electrons. To see smooth oscillations of the DOS we also average the calculated LDOS over different sites and realizations of uncorrelated vortices. In other words the quantity of our central interest is
\begin{equation}
\rho(B)= \left\langle\frac{1}{LM}\sum_{i=1}^{L/2} \sum_{j=1}^{2M} (-\frac{1}{\pi}) \mathrm{Im}\; G_{i}(j,j;0+i \,\delta)\right\rangle,
\end{equation}
where the angular brackets denote average over independent vortex realizations.
In the Green's function we have already set the energy to the chemical potential $E=0$. For all the numerical results presented in the following, an infinitesimal energy broadening $\delta=0.005 t$ will be chosen, if not specified otherwise, and the periodic boundary condition is imposed in the $y-$direction.

\section{Results}
\subsection{The Onsager rule for quantum oscillation frequencies}
\subsubsection{The two-fold DDW order case}
\begin{figure*}[ht]
  \centering
  \subfloat[FS for the two-fold DDW]
  {
  \includegraphics[trim=0mm 0mm 0mm 0mm,width=0.3\textwidth]{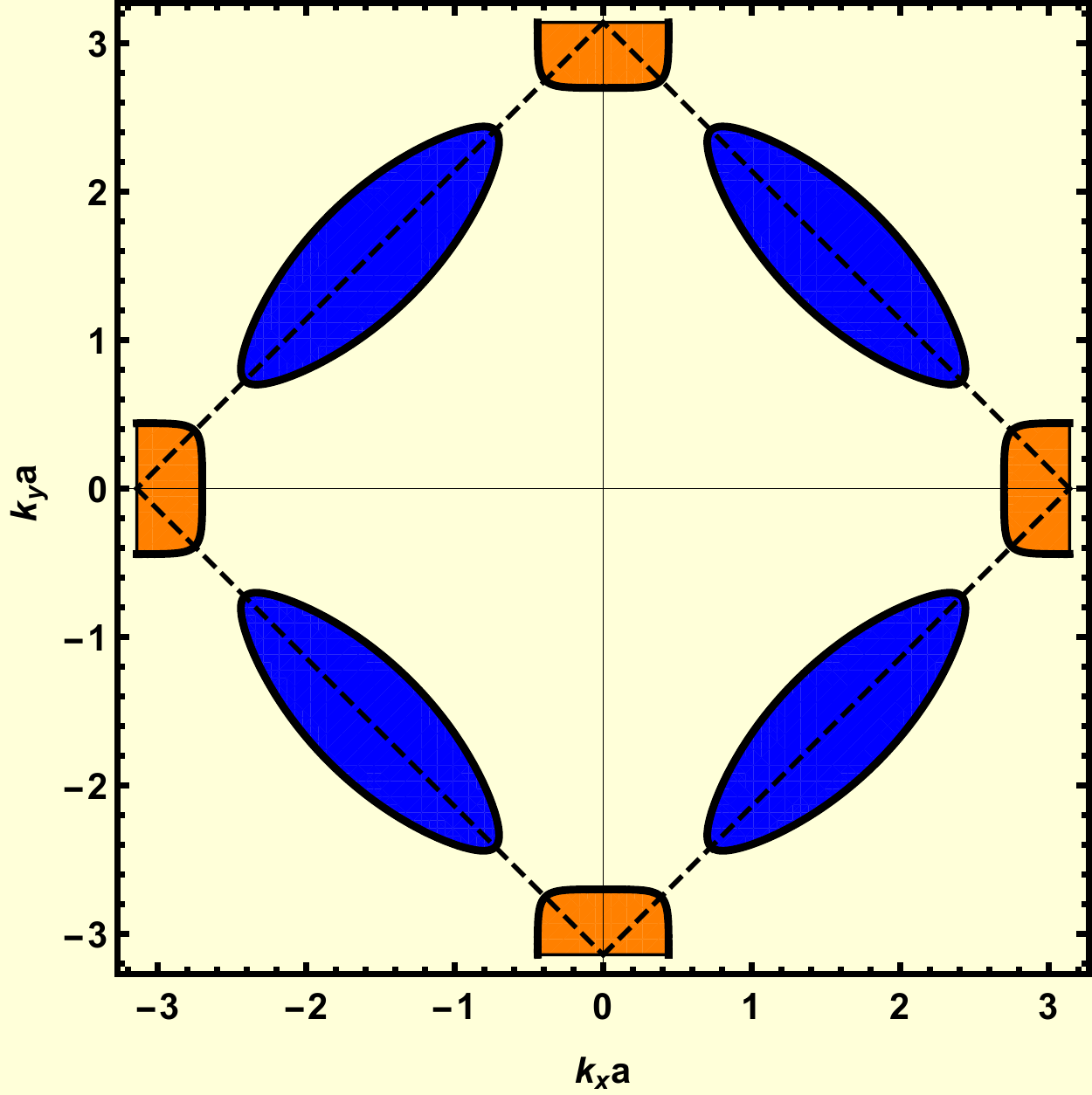}
   \label{fig:FS-2fddw}
  }
  \quad
  \centering
  \subfloat[FS for the bi-directional CDW]
  {
  \includegraphics[trim=0mm 0mm 0mm 0mm,width=0.3\textwidth]{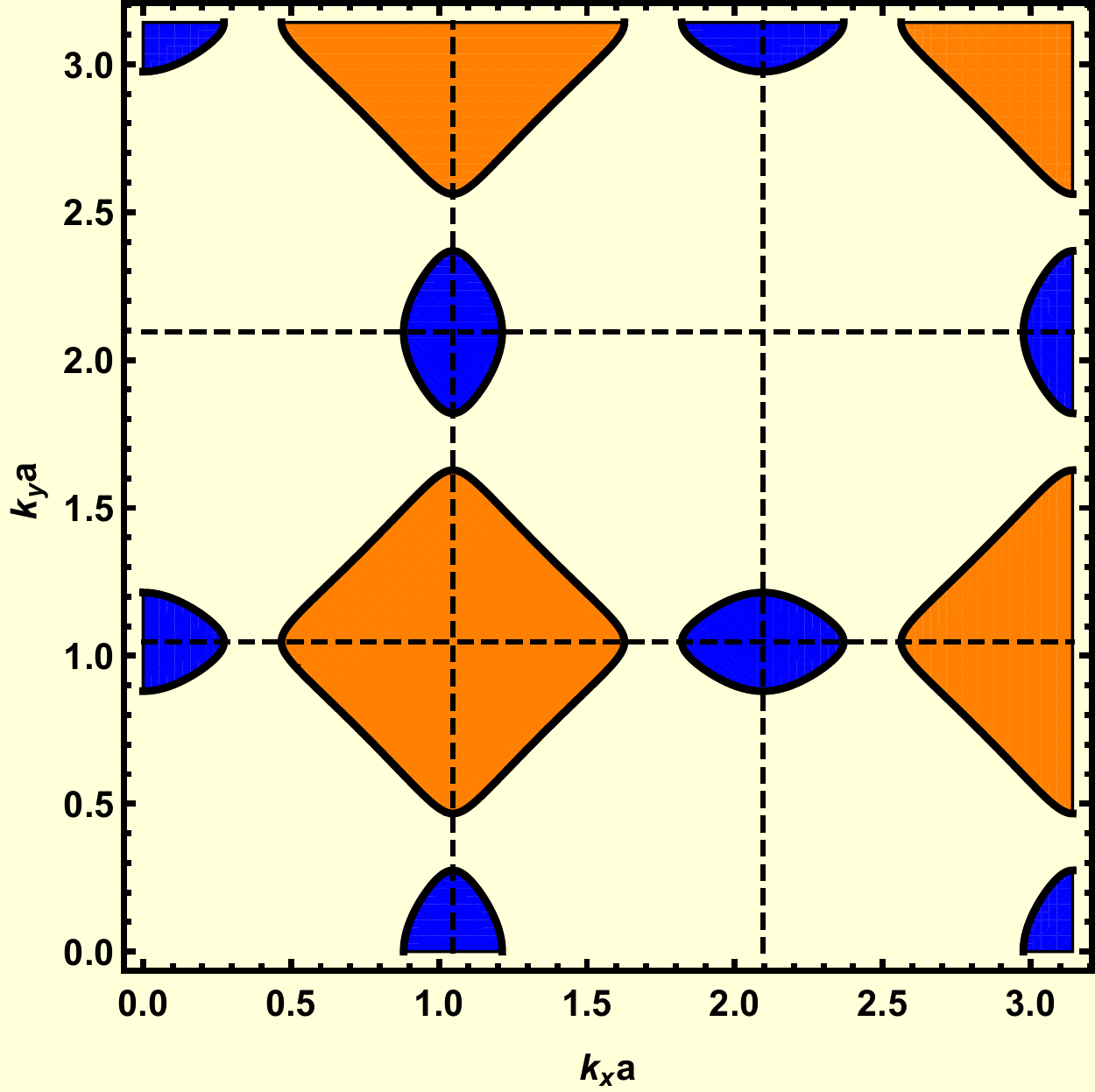} 
   \label{fig:FS-bcdw}
  }
  \quad
  \centering
  \subfloat[FS for the period-$8$ DDW]
  {
  \includegraphics[trim=0mm 0mm 0mm 0mm,width=0.3\textwidth]{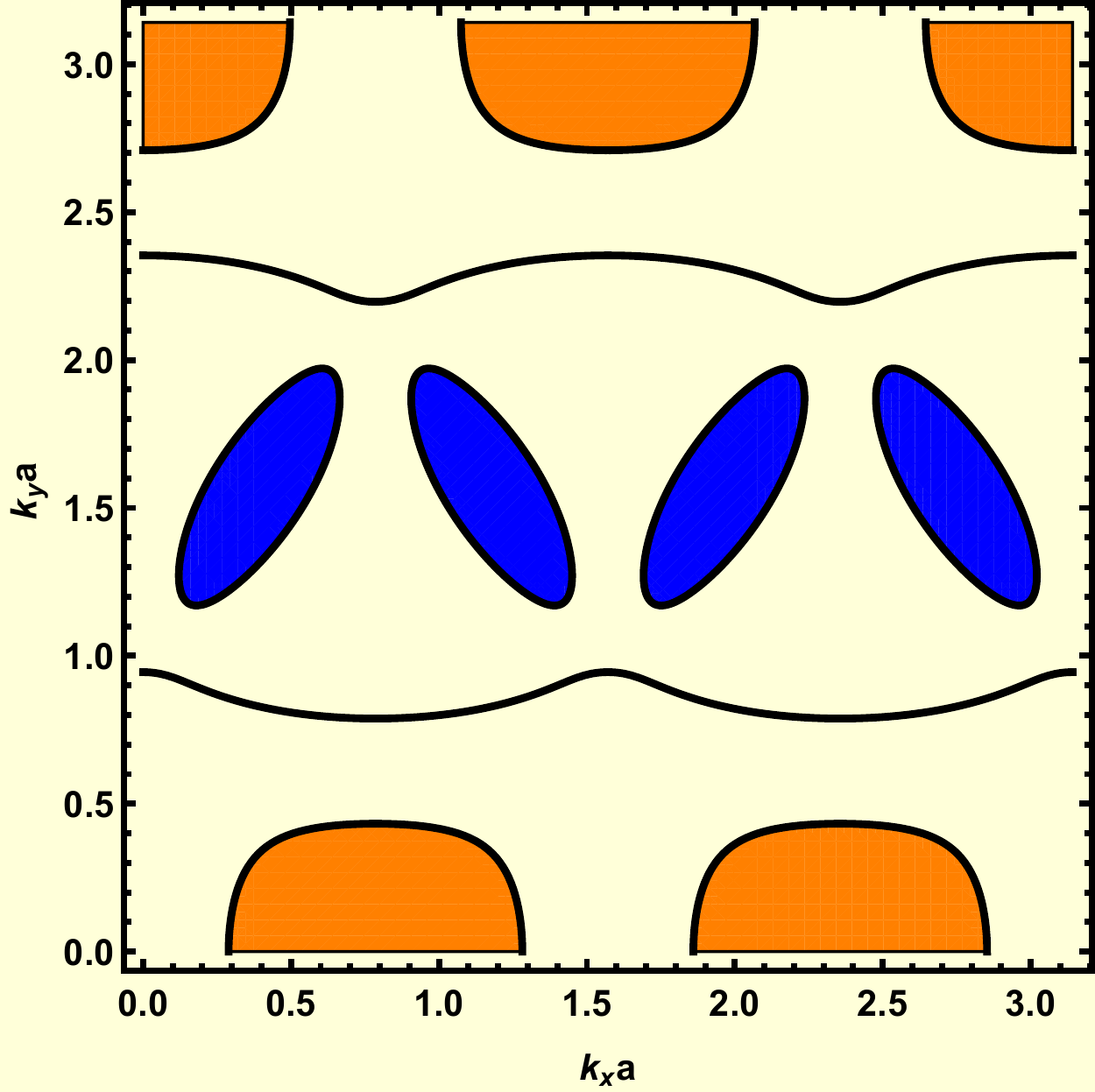}
   \label{fig:FS-p8ddw}
  }
  \newline
  \subfloat[FFT spectrum for the two-fold DDW]
  {
  \includegraphics[trim=0mm 0mm 0mm 0mm,width=0.3\textwidth]{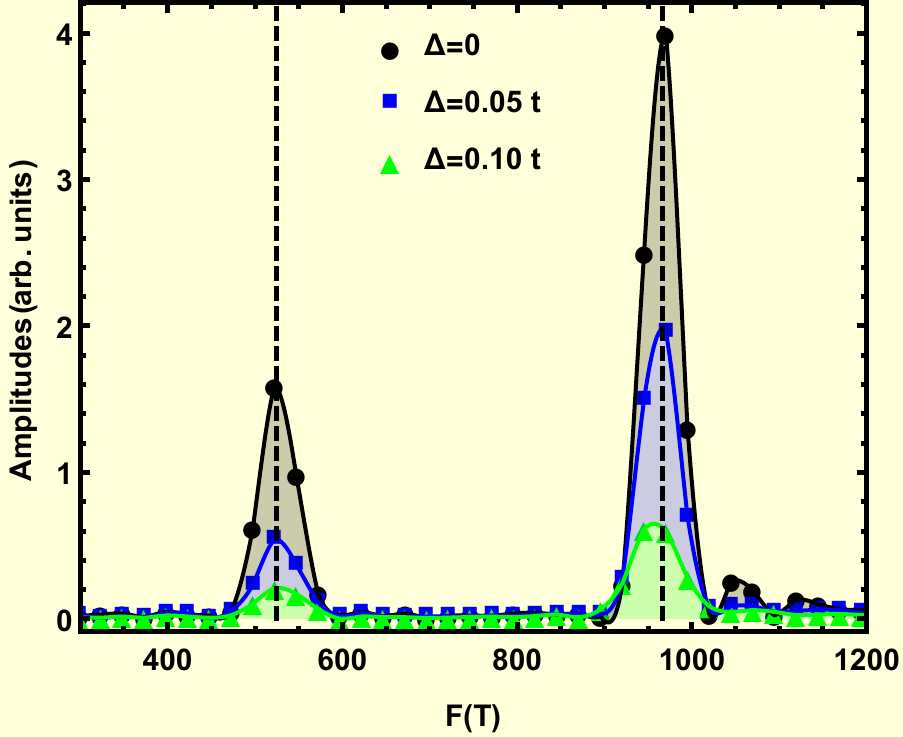}
   \label{fig:dos-t-tp-tpp-2fddw-vt-FFT-2}
  }
  \quad
  \subfloat[FFT spectrum for the bi-directional CDW]
  {
  \includegraphics[trim=0mm 0mm 0mm 0mm,width=0.3\textwidth]{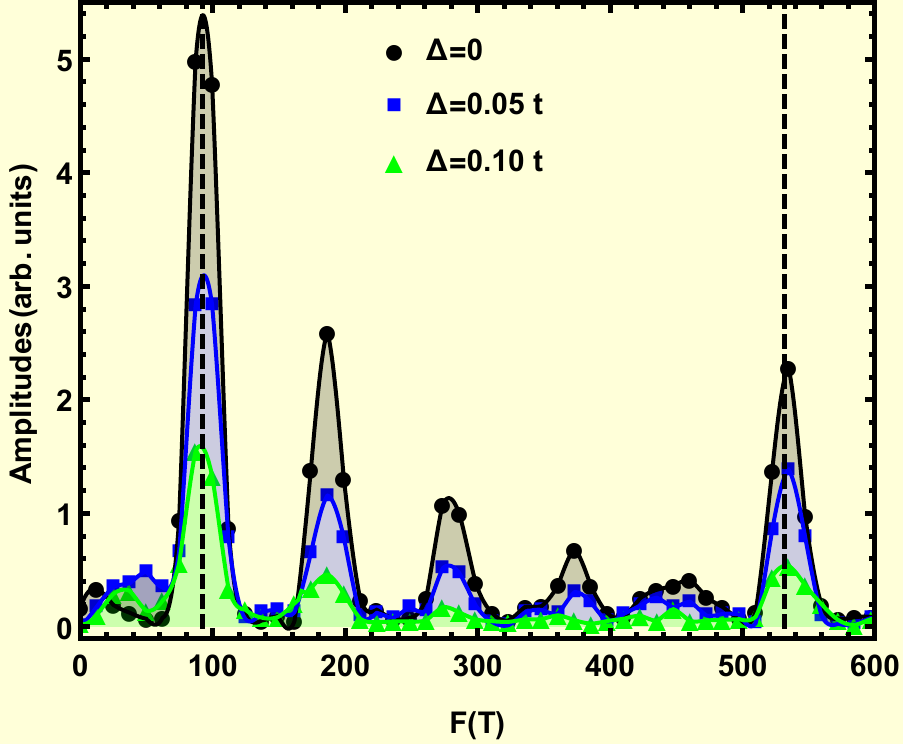}
   \label{fig:dos-tpp-bcdw-p2-FFT}
  }
  \quad
  \subfloat[FFT spectrum for the period-8 DDW]
  {
  \includegraphics[trim=0mm 0mm 0mm 0mm,width=0.3\textwidth]{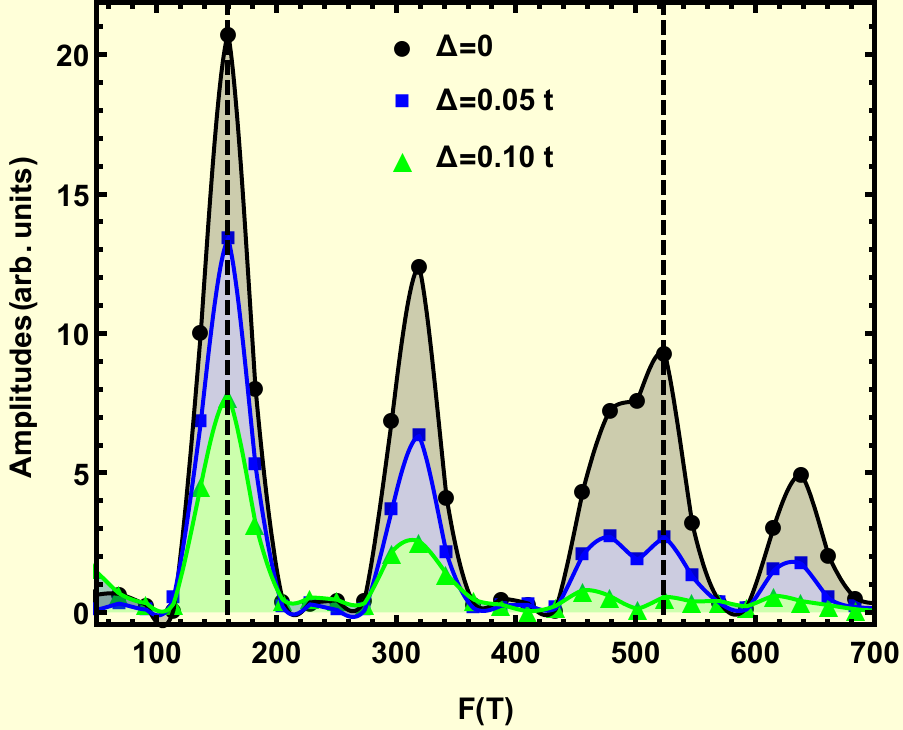}
   \label{fig:dos-tpp-p8ddw-FFT}
  }
  \centering
  \caption{The upper panel shows the plots of Fermi surfaces(FS), with electron pockets shaded in orange and hole pockets in blue. In Fig.~\ref{fig:FS-2fddw}, the area enclosed by the dashed lines gives the reduced Brillouin zone. In Fig.~\ref{fig:FS-bcdw}, the open orbits are not shown for clarity. And the dashed lines denote the positions $k_xa,k_ya=\pi/3,2\pi/3$. In the lower panel we show the corresponding oscillation FFT spectrums for various values of $\Delta$. The two vertical dashed lines in each plot denote the two fundamental oscillation frequencies calculated from the Fermi surface area via the Onsager relation. They are $F_{\mathrm{e}}= 525 \mathrm{T},F_{\mathrm{h}}= 966 \mathrm{T}$ in Fig.~\ref{fig:dos-t-tp-tpp-2fddw-vt-FFT-2}, $F_{\mathrm{e}}=529\mathrm{T},F_{\mathrm{h}}=92 \mathrm{T}$ in Fig.~\ref{fig:dos-tpp-bcdw-p2-FFT}, and $F_{\mathrm{e}}=523\mathrm{T},F_{\mathrm{h}}=159 \mathrm{T}$ in Fig.~\ref{fig:dos-tpp-p8ddw-FFT}. Other parameters used are $L=1000,M=100,\delta=0.005 t$ in Fig.~\ref{fig:dos-t-tp-tpp-2fddw-vt-FFT-2}, $L=1000,M=102,\delta=0.005t$ in Fig.~\ref{fig:dos-tpp-bcdw-p2-FFT}, and $L=1000,M=200,\delta=0.002t$ in Fig.~\ref{fig:dos-tpp-p8ddw-FFT}. }
  \label{fig:dos-FFT}
\end{figure*}

With the parameters: $t=1,t^{\prime}=0.30\, t, t^{\prime\prime}=t^{\prime}/9.0,\mu=-0.8807\,t,W_0=0.26\, t,V_c=0$, the hole doping level is $p\approx 11\%$. Without vortices we can diagonalize the Hamilontian $H$ in the momentum space and obtain the normal state Fermi surface. This Fermi surface consists of two closed orbits, see the inset of Fig.~\ref{fig:FS-2fddw}. The bigger one centered around the node point $(\frac{\pi}{2},\frac{\pi}{2})$ is hole like. It has an area $\frac{A_{\mathrm{h}}}{(2\pi/a)^2}\approx 3.47\%$. This corresponds to an oscillation frequency $F_{\mathrm{h}}= \frac{A_{h}}{(2\pi/a)^2} \frac{2\Phi_s}{a^2}= 966 \mathrm{T}$ from the Onsager relation, where $\Phi_s=hc/2e$ is the fundamental flux quanta and the two lattice spacings $a^2=3.82\AA\times 3.89\AA$ are chosen for $\mathrm{YBCO}$. At the antinodal point $(0,\pi)$ there is an electron pocket with an area $\frac{A_{\mathrm{e}}}{(2\pi/a)^2}\approx 1.9\%$, corresponding to a frequency $F_{\mathrm{e}}=525 \mathrm{T}$ (electron).  We should notice that the fast oscillation $F_{\mathrm{h}}$ (hole) is not observed in the experiments in cuprates. This problem can be resolved if we consider a period$-8$ DDW model~\cite{Eun2012}; see below.

We compute the $\rho(B)$ as a function of the inverse of the magnetic field  $1/B$ in the presence of various  $\Delta$. In these calculations, the number of vortices are chosen such that the total magnetic flux is equal to $\Phi=B LMa^2$. From the oscillatory part of  $\rho(B)$ we  perform Fast Fourier Transform(FFT) to get the spectrum. The result is shown in Fig.~\ref{fig:dos-t-tp-tpp-2fddw-vt-FFT-2}. In this spectrum the two oscillation frequency $F_{\mathrm{e}}= 525 \mathrm{T}$ and $F_{\mathrm{h}}= 966 \mathrm{T}$  calculated from the normal state Fermi surface areas via the Onsager relation are also shown by the two vertical dashed lines. We see clearly that as we increase  $\Delta$ the oscillation amplitudes are damped. However, remarkably, the oscillation frequencies remain the same within numerical errors. Thus, even in the presence of vortices, the Onsager rule still holds to an excellent approximation.


\subsubsection{The bi-directional CDW order case}
We choose the following parameters: $t=1,t^{\prime}=0.2 t,t^{\prime\prime}=t^{\prime}/8,V_c=0.12t,\mu=-0.73 t$ so that we can produce the right oscillation frequencies that are observed in experiments. The hole doping level is $p\approx 11\%$. The Fermi surface of the normal state is plotted in the inset of Fig.~\ref{fig:FS-bcdw} (open orbits are not shown for clarity). There are two closed Fermi surface sheets. Centered around the point $(\frac{\pi}{3},\frac{\pi}{3})$ and other symmetry related positions there are diamond shaped electron pockets, highlighted in orange. This pocket has an area $\frac{A_{\mathrm{e}}}{(2\pi/a)^2}=1.9\%$. It corresponds to a frequency $F_{\mathrm{e}}=529\mathrm{T}$ from the Onsager relation. Besides this electron pocket, there is an oval shaped hole pocket centered around $(\frac{\pi}{3},\frac{2\pi}{3})$, highlighted in blue. The area of this hole pocket is $\frac{A_{h}}{(2\pi/a)^2}=0.33\%$. This corresponds to an oscillation frequency $F_{\mathrm{h}}= 92 \mathrm{T}$.

The oscillation spectrum of the $\rho(B)$ is shown in  Fig.~\ref{fig:dos-tpp-bcdw-p2-FFT}. From the spectrum we see that when the vortex scattering is absent, $\Delta=0$, the oscillation amplitudes peak at the two frequencies $F_{\mathrm{e}},F_{\mathrm{h}}$, as denoted by the two vertical dashed lines. These results agree with our Fermi surface calculation, as we expected.  When the vortices are included the oscillation amplitude is gradually damped as the vortex scattering strength is increased by increasing $\Delta$. However whenever the oscillation frequency can be clearly resolved, we see that their positions do not change with $\Delta$. Again this means that the Onsager rule survives in the presence of vortex scattering.


\subsubsection{The period$-8$ DDW order case}\label{sec:p8ddw}
In this subsection we present our quantum oscillation results for the period$-8$ DDW model. In this model the period$-8$ stripe DDW order is considered as the major driving force behind the Fermi surface reconstructions; while a much weaker unidirectional period$-4$ CDW is included as a subsidiary order.

We choose the parameter set $t^{\prime}=0.3 t, t^{\prime\prime}=t^{\prime}/2.0,  W_0=0.70 t, V_c=0.05 t,\mu=-0.70 t$ and estimate the hole doping level to be $p\approx 11.2\%$. We also obtain a Fermi surface similar to the one we had in Ref.\cite{Eun2012}. It has a large pocket of electron like with a frequency $F_{\mathrm{e}}=523\mathrm{T}$, a smaller pocket of hole like with a frequency $F_{\mathrm{h}}=159\mathrm{T}$, and also some open orbits which do not contribute to quantum oscillations.

The corresponding oscillation spectrum is presented in Fig.~\ref{fig:dos-tpp-p8ddw-FFT}, where we see the oscillation amplitude decreases as we increase $\Delta$, however, the frequencies do not change with $\Delta$. In other words the presence of vortex scattering does not alter the oscillation frequencies.

The observations here, combined with the other two cases, strongly suggest that the Onsager's relation being intact in the presence of vortex scattering is generic and independent of the order parameters that reconstruct the Fermi surface.


\subsection{The Density of states at high fields}
In the above we have examined the effects of random vortex scattering on the quantum oscillations. Now we give an overview of the $B$ dependence of the DOS for fields $B \gtrsim 10\mathrm{T}$, at a representative value of $\Delta=0.1 t$. At lower fields, the vortex liquid model is not valid any more, since vortices should order into a solid instead. Therefore we should use a vortex lattice to model such a state. In the following we focus on the high field regime first and defer our vortex lattice discussions for the low field regime to the later section~\ref{sec:vtx-solid}.

\subsubsection{The period-2 DDW order case}
\begin{figure*}[ht]
  \centering
  \subfloat[The two-fold DDW order]
  {
  \centering
  \includegraphics[trim=0mm 0mm 0mm 0mm,width=0.3\textwidth]{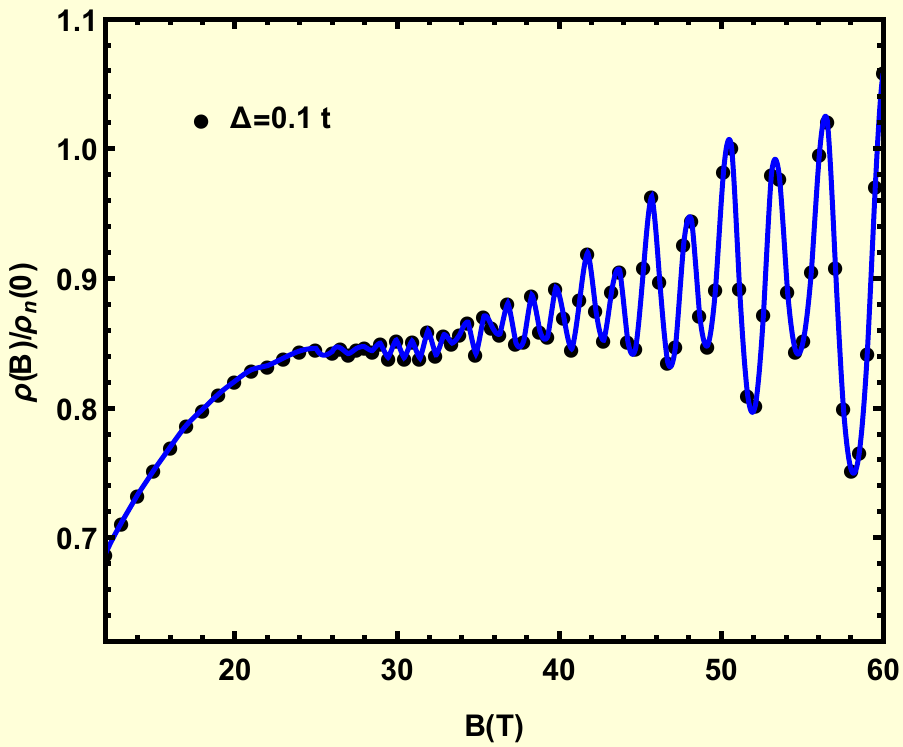}
   \label{fig:dos-tpp-2fddw-overview}
  }
  \quad
  \subfloat[The bi-directional CDW order]
  {
  \centering
  \includegraphics[trim=0mm 0mm 0mm 0mm,width=0.3\textwidth]{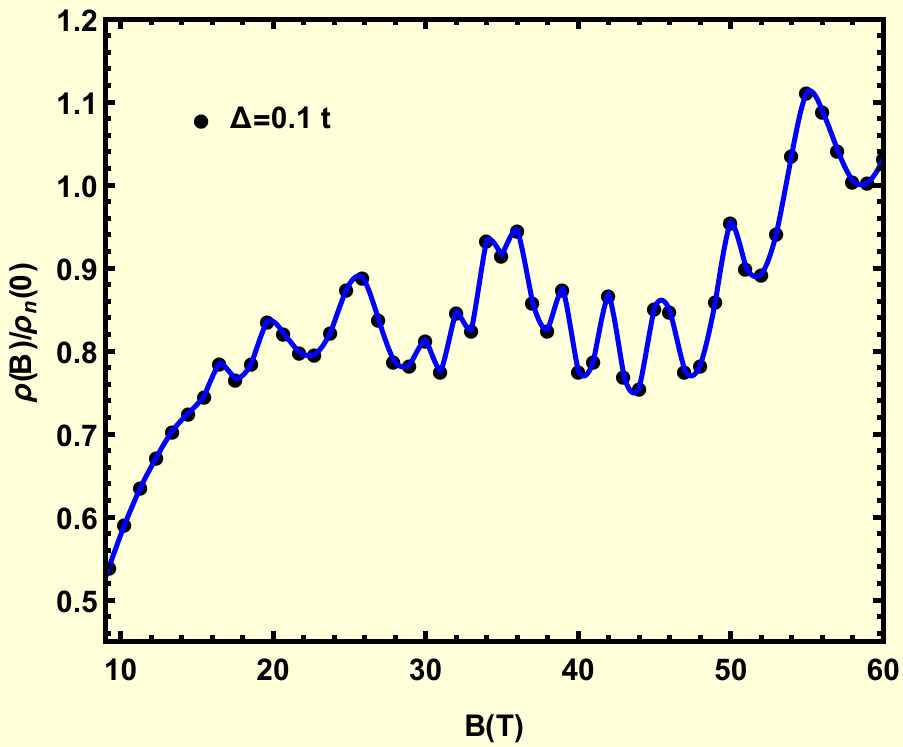}
   \label{fig:dos-bcdw-overview-d10-p2}
  }
  \quad
   \subfloat[The period-8 DDW order]
  {
  \centering
  \includegraphics[trim=0mm 0mm 0mm 0mm,width=0.3\textwidth]{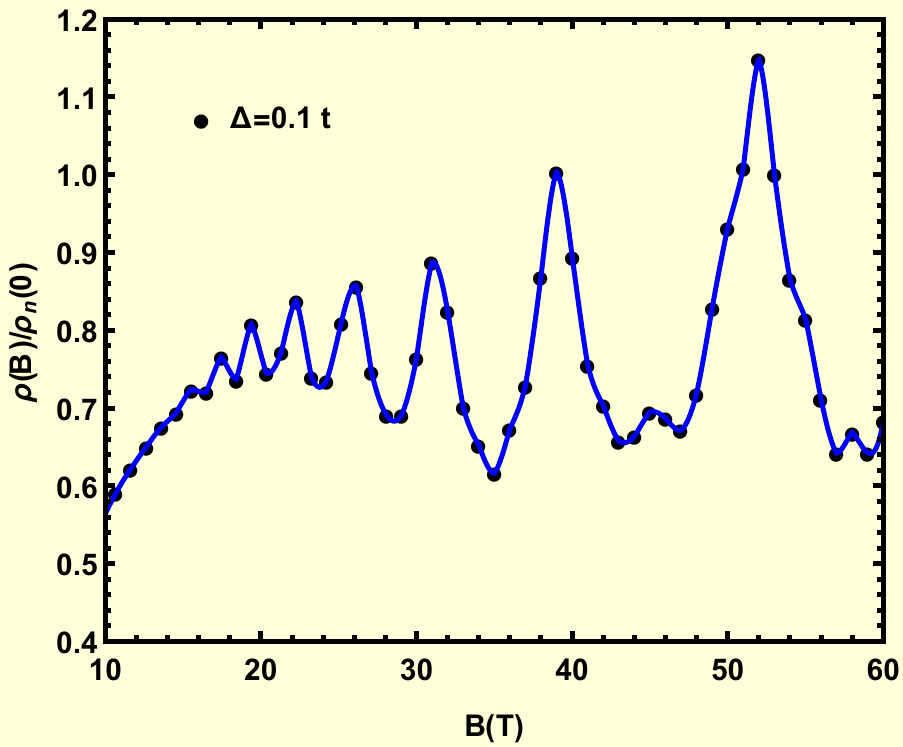}
   \label{fig:dos-p8ddw-d10-overview}
  }
  \centering
  \caption{The DOS $\rho(B)$,normalized to the normal state DOS $\rho_{\mathrm{n}}(0)$ at zero field $B=0$ for different cases. The estimated values of the normal state DOS are $\rho_{\mathrm{n}}(0)\approx 0.23 \,\mathrm{ states/t}$ for the two-fold DDW, $\rho_{\mathrm{n}}(0)\approx 0.25 \; \mathrm{states}/t$ for the bi-directional CDW, and $\rho_{\mathrm{n}}(0)\approx 0.18 \; \mathrm{states}/t$ for the period-$8$ DDW case. The data is averaged over $108$, $120$,$40$ different vortices configuration realizations respectively. }
  \label{fig:dos-d10-overview}
\end{figure*}

In Fig.~\ref{fig:dos-tpp-2fddw-overview} we plot $\rho(B)/\rho_n(0)$ as a function of the field $B$ for the two-fold DDW order case, where $\rho_{\mathrm{n}}(0)$ is the normal state DOS at zero field. In the following,  the normal state should be understood as a state, which does not have any superconductivity but can have a particle-hole density wave order. And all the DOS value calculated is for one electron in a single $\mathrm{CuO}$ plane, without including the spin degeneracy. From Fig.~\ref{fig:dos-tpp-2fddw-overview} we see that as $B$ decreases, the DOS oscillation gets suppressed gradually. This is because the orbital quantization of electrons becomes dominated  by the vortex scattering.

A noticeable feature of this plot is that when the field becomes large, the oscillation of $\rho(B)$ in $1/B$ gradually develops on top of a constant background. This constant background value of $\rho(B)$ is suppressed from the normal state DOS $\rho_{\mathrm{n}}(0)$. The size of this suppression depends on the vortex scattering strength. For the parameters used in Fig.~~\ref{fig:dos-tpp-2fddw-overview} it is $\sim 15\%$. This constant background of $\rho(B)$ is different from the previous results obtained in  Fig. 3(b) of the Ref.~\cite{Banerjee2013} in the absence of vortices.

\begin{figure}[htp]
  \centering
  \includegraphics[width=0.45\textwidth]{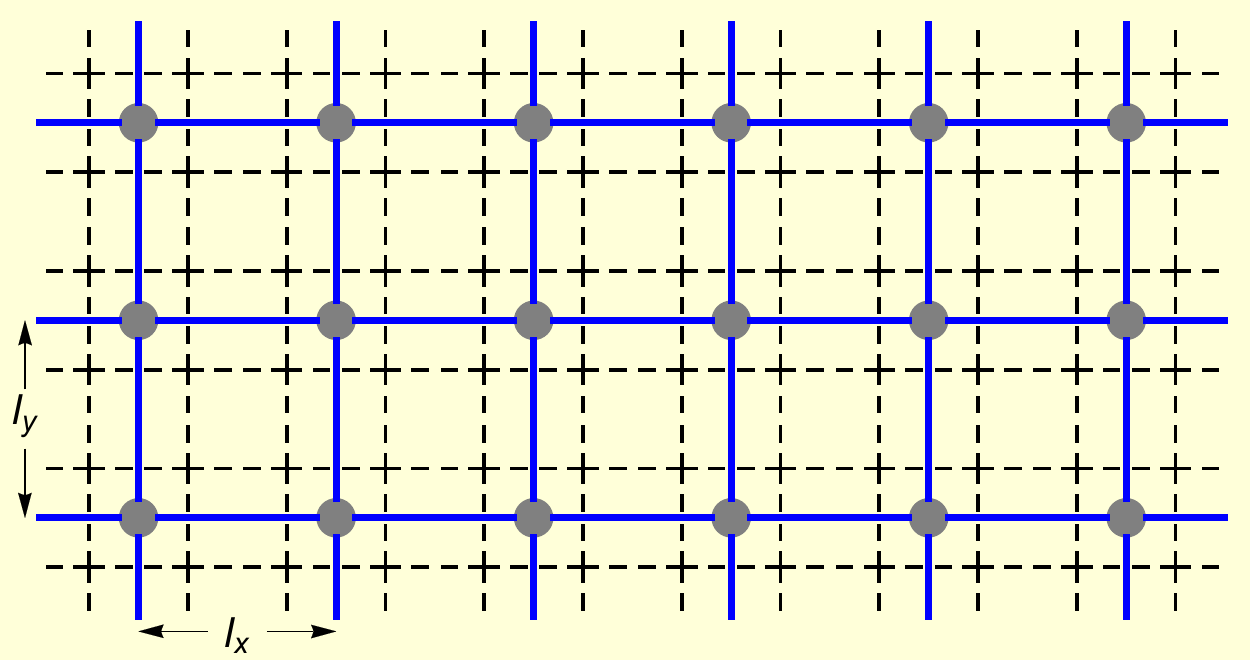}
  \caption{Schematic diagram of a square vortex lattice. The dashed lines represent the original $\mathrm{CuO}$ lattice; while the full lines stand for the vortex lattice, with each vortex, represented by the grey disks, sitting at the $\mathrm{CuO}$ plaquette center. And $(l_x,l_y)$ are the vortex lattice spacings, in units of the original $\mathrm{CuO}$ square lattice spacing $a$.}
  \label{fig:vtxsolid}
\end{figure}

\begin{figure*}[ht]
  \centering
  \subfloat[The pure d-wave vortex lattice case without any other density wave order at the optimal hole doping]
  {
  \includegraphics[trim=0mm 0mm 0mm 0mm,width=0.4\textwidth]{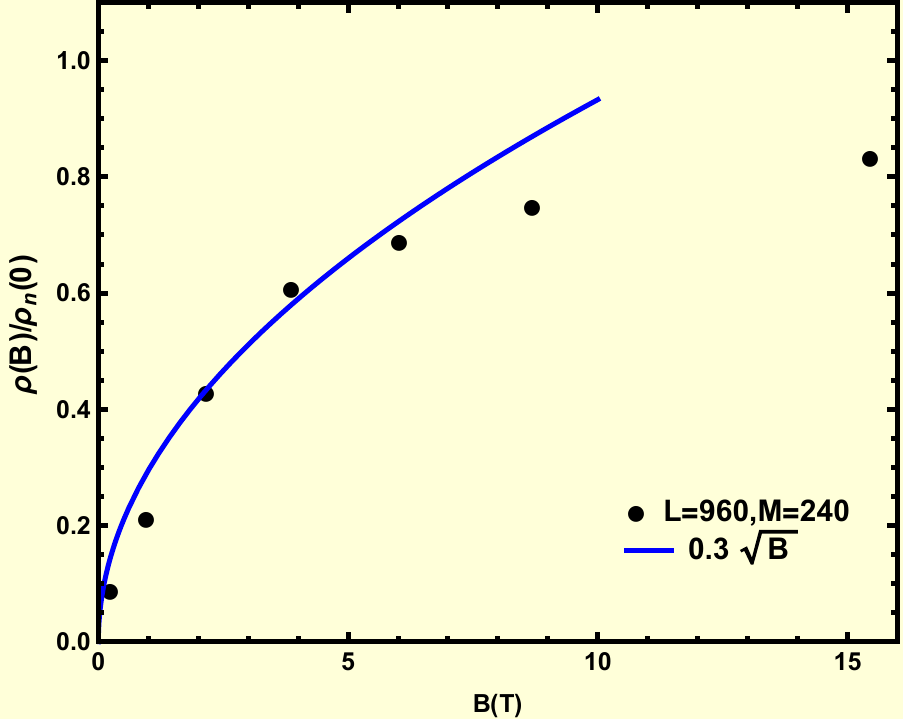}
   \label{fig:D01t-pure-vtxs-DoS}
  } 
  \qquad \qquad
  \subfloat[The coexistence of a vortex lattice and a two fold DDW order in the underdoped regime]
  {
  \includegraphics[trim=0mm 0mm 0mm 0mm,width=0.4\textwidth]{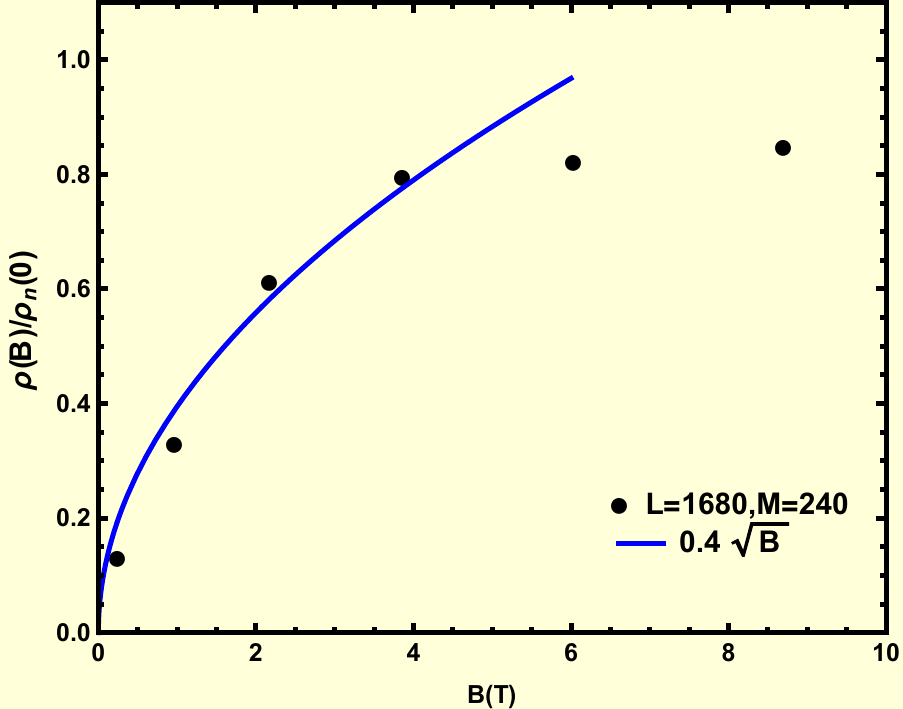}
   \label{fig:D01t-2fddw-vtxs-DoS}
  }
  \centering
  \caption{The DOS of vortex solids for $\Delta=0.1 t$. In Fig.~\ref{fig:D01t-pure-vtxs-DoS}, $\rho_n(0) \approx 0.25 \; \mathrm{ states}/t$ and in Fig.~\ref{fig:D01t-2fddw-vtxs-DoS} $\rho_n(0) \approx 0.23 \; \mathrm{ states}/t$. }
  \label{fig:dos-vtxsolid}
\end{figure*}


\subsubsection{The bi-directional CDW order case}
The constant background of the DOS oscillation is not restricted to the two-fold DDW order case. As we can see in Fig.~\ref{fig:dos-bcdw-overview-d10-p2}, for the bi-directional CDW order, the $\rho(B)$ oscillation background is again a constant at high fields.

\subsubsection{The period$-8$ DDW order case}
We also confirm this constant $\rho(B)$ background feature in the oscillation regime for the period$-8$ DDW order case in Fig.~\ref{fig:dos-p8ddw-d10-overview}.

Therefore we can conclude that the high field $\rho(B)$ oscillation background being a constant is  generic.


\subsection{Vortex solid at low fields}\label{sec:vtx-solid}
Now we move on to the low field regime. In this regime when the field is low enough,  the vortices order into a lattice. Whether the lattice is square  or triangular requires a self-consistent computation of the system's free energy, which is
far beyond the scope of this paper. Instead we simply take a square lattice for illustration. But none of the following qualitative features should depend on the vortex lattice type.

\subsubsection{Implementation of the square vortex lattice}
To put the square vortex lattice onto our original $\mathrm{CuO}$ lattice so that each vortex sits at the $\mathrm{CuO}$ lattice plaquette center and the periodic boundary condition is still preserved along the transverse direction, we require the vortex lattice to be commensurate with our original $\mathrm{CuO}$ lattice, as schematically shown in Fig.~\ref{fig:vtxsolid}. Namely, if the vortex lattice spacings are $(l_x,l_y)$, and the corresponding vortex lattice size is $(N_x,N_y)$, we require that the original $\mathrm{CuO}$ lattice size $(L,M)$ satisfies $L=N_x \, l_x, M=N_y\, l_y$. For a particular value of $(L,M)$, this restricts the possible values of $(l_x,l_y)$ and also the possible values of the magnetic field, because the vortex lattice spacings $(l_x ,l_y )$ are connected to the magnetic flux density via $ B=\Phi_s/(l_x l_y a^2)$, where $\Phi_s=hc/2e$ is the fundamental flux quanta. In our following calculation we pick a particular value of the system size $(L,M)$, find all the possible compatible values of the vortex lattice spacings $l_x=l_y$, and then for each of them calculate the magnetic field $B$ as well as the corresponding DOS.

However, we should calculate the DOS of the Bogoliubov quasiparticles instead of the electrons, because the system is far from being in a normal state in such a low field regime. Therefore now $\rho(B)$ is computed from the following formula instead
\begin{equation}
\rho(B)= \frac{1}{2} \frac{1}{LM}\sum_{i=1}^{L/2} \sum_{j=1}^{4 M} (-\frac{1}{\pi}) \mathrm{Im}\; G_{i}(j,j;0+i \,\delta).
\end{equation}
The major differences here from the one we used in our quantum oscillation calculations are: (1) the summation of the Green's function's diagonal matrix elements includes both the electron part and the hole part: $j$ runs from $j=1$ to $j=4M$ instead of $j=2M$; (2) there is no averaging over different vortices configurations because the vortex lattice is ordered; (3) an additional prefactor of $1/2$ is added to avoid double counting of degrees of freedoms.

For such a vortex lattice calculation, the summation over different vortex positions in the superfluid velocity calculation in Eq.~\eqref{eq:vs2} can be done exactly by using
\begin{gather}
\sum_n e^{i \vec{k}\cdot(\vec{r}-\vec{R}_n)}=N_x N_y \, \sum_{(n_1,n_2)} e^{i\vec{G}_{n_1,n_2}\cdot\vec{r}},
\end{gather}
where $\vec{G}_{n_1,n_2}=( \frac{2 n_1 \pi}{l_x a}, \frac{2 n_2 \pi}{l_y a})$ is a reciprocal Bragg vector of the square vortex lattice, with $n_1, n_2 \in \mathbb{Z}$. Then the Eq.~\eqref{eq:vs2} of $\vec{v}_s$ becomes
\begin{gather}
m \vec{v}_s  = \pi \hbar \frac{1}{l_x l_y a^2} {\sum_{(n_1,n_2)}}^{\prime} \frac{\vec{G}_{n_1,n_2}\times \hat{z}}{|\vec{G}_{n_1,n_2}|^2} \sin[\vec{G}_{n_1,n_2}\cdot \vec{r}]
\end{gather}
The summations of $(n_1,n_2)$ are restricted to those values that satisfy $ 0 \le \frac{2n_1\pi}{l_x a} < \frac{2\pi}{a}, 0 \le \frac{2 n_2 \pi}{l_y a} < \frac{2\pi}{a} $. Again the prime superscript in the summation means the point $(n_1,n_2)=(0,0)$ is excluded.

\subsubsection{DOS numerical results}
 According to Volovik~\cite{Volovik1993}, for a $d_{x^2-y^2}-$wave vortex, the major contribution to the low energy DOS comes from the extended states along the  nodal direction. In his semiclassical analysis this contribution is computed from the Doppler shift  of the quasiparticle energy. The conclusion is that the DOS for a single vortex is $\rho(B)\propto 1/\sqrt{B}$. In the limit that the number of vortices is proportional to $B$, which is not valid if $B$ is near the lower critical field $H_{c1}$, multiplying it by the number of vortices gives $\rho(B)\propto \sqrt{B}$. Extrapolating this result to the high field regime and using the fact that near the upper critical field $H_{c2}$, $\rho(B)$ should roughly recover the normal state DOS $\rho_n(0)$, he concluded that $\rho(B)/\rho_n(0)=\kappa \sqrt{B/H_{c2}}$, with $\kappa$ some constant of order unity. This type of analysis is applicable only in the small field limit in the sense that $B \ll H_{c2}$ so that each vortex is far apart from any others. This is exactly the field regime where the vortex solid state develops. In the following we compute the DOS for a $d-$wave vortex lattice, for the cases both with and without an additional particle-hole density wave order, and test them against Volovik's results. For our following comparisons we slightly rewrite the above field dependence of $\rho(B)$ as follows
\begin{gather}
\frac{\rho(B)}{\rho_n(0)}=\kappa \frac{\sqrt{B}}{\sqrt{H_{c2}}}= \kappa \sqrt{\frac{2\pi \xi^2}{\Phi_s}} \sqrt{B}\approx 0.1 \kappa \sqrt{B}, \label{eq:volovik}
\end{gather}
where $\frac{\Phi_s}{2\pi\xi^2}\approx 90\mathrm{T}$, if $\xi=5 a$ and $a\approx 3.83\AA$ are used.

\begin{enumerate}
\item First we consider a square vortex lattice without any other additional density wave order. We choose the band structure parameters to be $t^{\prime}/t=0.3,t^{\prime\prime}=t^{\prime}/9.0,\mu=-1.01 t$ so that the estimated normal state hole doping level $p\approx 15\%$ is at the optimal doping. The computed DOS is shown in Fig.~\ref{fig:D01t-pure-vtxs-DoS}. At low enough fields, all the data points follow the $\rho(B)/\rho_n(0)=0.3 \sqrt{B}$ line, although there is some small scatter in the data, which comes from the finite size effects of our vortex lattice. This $0.3\sqrt{B}$ corresponds to $\kappa\approx 3$ in Eq.~\eqref{eq:volovik}. To make a comparison with the specific heat measurements on $\mathrm{YBCO123}$ at the optimal doping~\cite{Moler1997}, we estimate the field dependent electronic specific heat $\gamma(B)$ from our DOS $\rho(B)$ as follows
\begin{gather}
\frac{\gamma(B)}{\gamma_n} = \frac{\rho(B)}{\rho_n(0)}\approx 0.1 \kappa \sqrt{B}.
\end{gather}
The normal state specific heat can be estimated as $\gamma_n=4\, \frac{\pi^2}{3}k_B^2 \rho_n(0)$. Here the additional prefactor of $4$ comes from the spin degeneracy and the fact that one unit cell of $\mathrm{YBCO123}$ contains two $\mathrm{CuO}$ planes. If we take $t=0.15 \mathrm{eV}$, then $\gamma_n\approx 15.7 \mathrm{mJ/mol\cdot K^2}$ and $\gamma(B)=A \sqrt{B}$ with the coefficient $A \approx 4.7 \, \mathrm{mJ/mol\cdot K^2\cdot T^{1/2}}$. Compared with the experimental value of $A\approx 0.9 \, \mathrm{mJ/mol\cdot K^2\cdot T^{1/2}}$ from Ref.~\cite{Moler1997}, our numerical value is greater by a factor of about $5$. This quantitative discrepancy is not significant given our approximations. In fact, it is quite reasonably consistent.

\item Next we consider the coexistence of a square vortex lattice and an additional two-fold DDW order in the underdoped regime. The parameters are the same as those in our high field quantum oscillation calculations: $t^{\prime}/t=0.3,t^{\prime\prime}=t^{\prime}/9.0,\mu=-0.8807 t, W_0=0.26 t$, so the estimated normal state, with the DDW order but no superconductivity, hole doping level is $p\approx 11\%$. Fig.~\ref{fig:D01t-2fddw-vtxs-DoS} shows the corresponding DOS results. The small field data follows $\rho(B)/\rho_n(0)=0.4\sqrt{B}$, corresponding to a value of $\kappa\approx 4$ in Eq.~\eqref{eq:volovik}. 
\end{enumerate}
 The above two values of $\kappa$ are consistent with the fact that in Volovik's formula $\kappa$ is of order unity. Of course its precise value depends on the vortex lattice structure, on the slope of the gap near the gap node (in the current case both the parameters $\Delta$ and $q$), and also on the normal state band structure.

 From the above two scenarios we can conclude that irrespective of the existence of an additional density wave order, the DOS of a clean vortex lattice always scales as $\rho(B)\propto \sqrt{B}$ in the low field limit.

\section{Conclusion}
In summary we have shown that in the quenched vortex liquid state the quantum oscillations in cuprates can survive at large magnetic fields. Although the oscillation amplitude can be heavily damped if the vortex scattering is strong,  the oscillation frequency is given by the Onsager rule to an excellent approximation. Of course, when the field is small the quantum oscillations are destroyed by the vortices and $\rho(B)$ gets heavily suppressed due to the formation of Bogoliubov quasiparticles. When the field is small enough, a vortex solid state forms instead and it can be modeled by an ordered vortex lattice. We show the field dependence of the vortex lattice's density of states follows $\rho(B)\propto \sqrt{B}$ in the asymptotically low field limit, in agreement with Volovik's semiclassical predictions. However in contrast to the previous suggestion our results show that this small field limit does not extend to the high field oscillatory regime of the vortex liquid state. Instead when the oscillations can be resolved, the non-oscillatory background of $\rho(B)$ flattens out, and becomes  field independent consistent with the more recent specific heat measurements~\cite{Marcenat2015}.

\begin{acknowledgments}
This research was supported by funds from David S. Saxon Presidential Term Chair at UCLA. We thank P. A. Lee for suggesting that we should explicitly check the results in  Ref.~\cite{Chen2009}. These  are shown in the Appendix~\ref{sec:patrick}. We used the Hoffman2 Shared Cluster provided by the UCLA Institute for Digital Research and Education program. We also thank Brad Ramshaw for discussion.
\end{acknowledgments}

\appendix
\section{Bond phase field $\theta_{ij}$ of  $\Delta_{ij}$}\label{sec:thetaij}
\begin{figure}[ht]
  \centering
  \subfloat[Bond $\overline{ij}$ does not cross the branch cut: $|\phi_i-\phi_j|<\pi$]
  {
  \includegraphics[trim=0mm 0mm 0mm 0mm,scale=0.3]{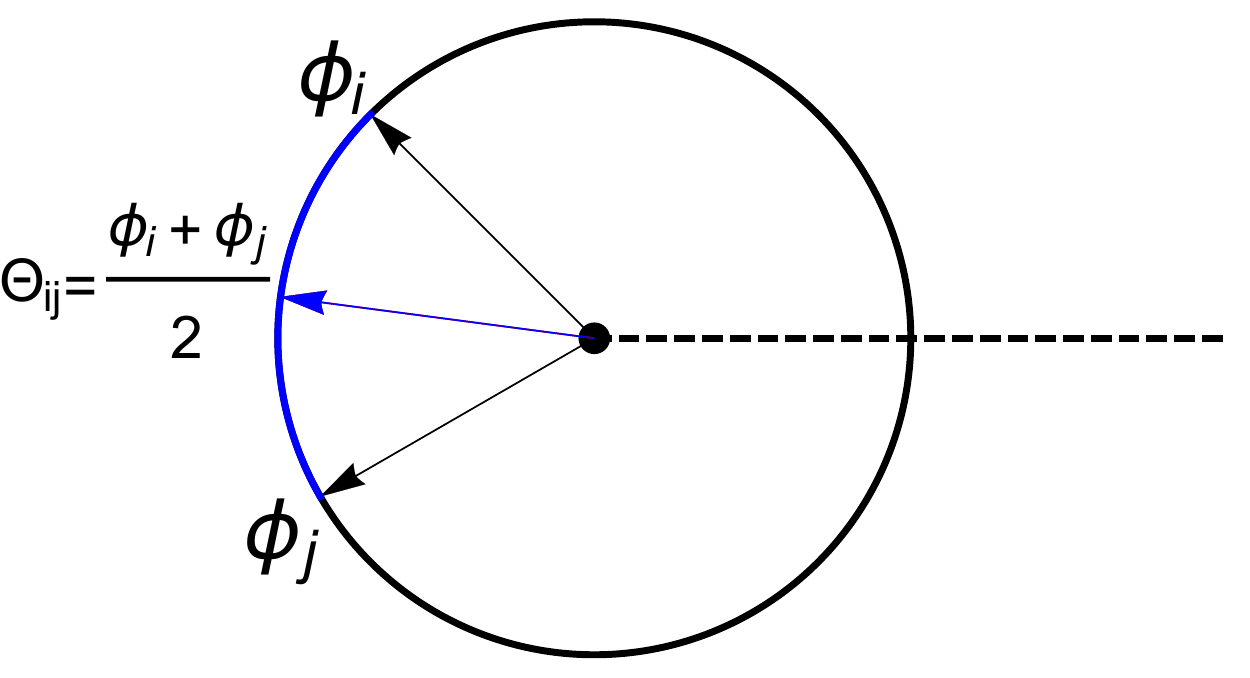}
   \label{fig:theta1}
  }
  \subfloat[Bond $\overline{ij}$ crosses the branch cut: $|\phi_i-\phi_j|>\pi$]
  {
  \includegraphics[trim=-10mm 0mm 0mm 0mm,scale=0.3]{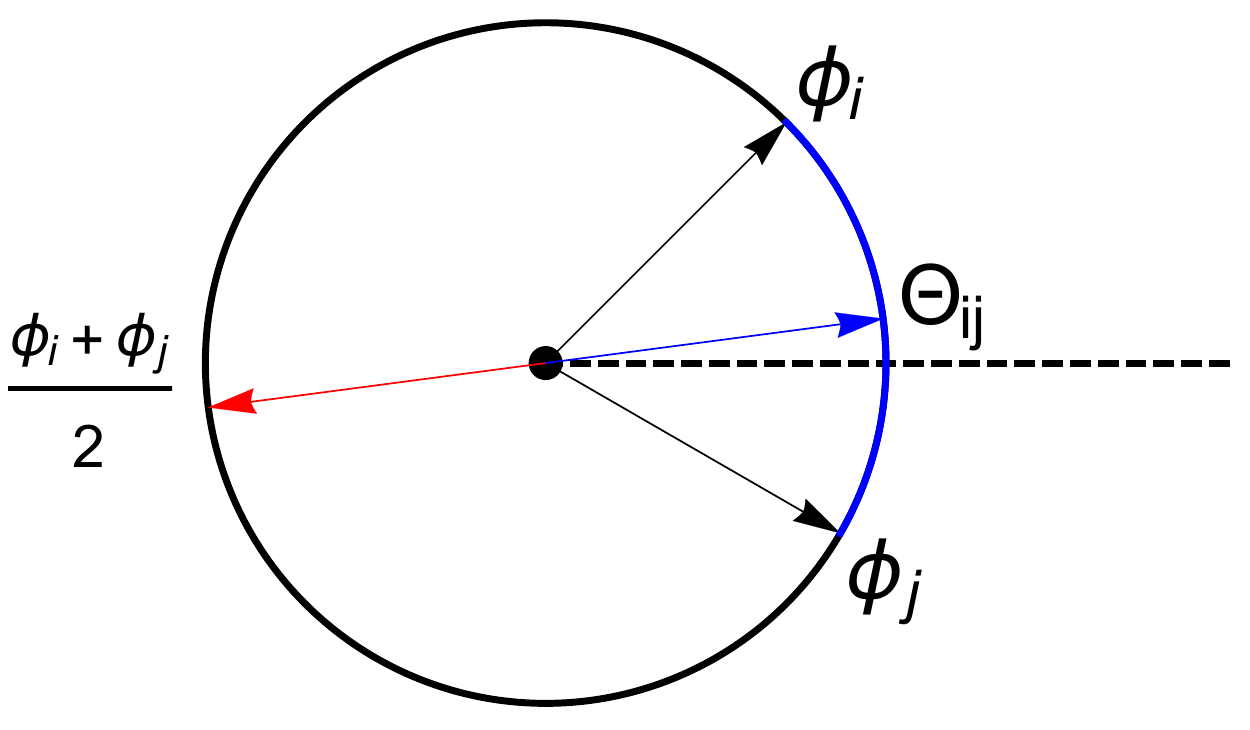}
   \label{fig:theta2}
  }
  \centering
  \caption{The black dot in the center represents a vortex. The dashed line, extended to the infinity, is its branch cut. The blue arc segment corresponds to the bond $\overline{ij}$ on a closed path. The phases $\phi_i,\phi_j$ are measured counter-clock wisely from the upper side of the branch cut. In Fig.~\ref{fig:theta1}, the bond $\overline{ij}$ does not cross the branch cut, and $(\phi_i+\phi_j)/2$ is a good definition for $\theta_{ij}$ ; however, if the bond $\overline{ij}$ crosses a branch cut, as in Fig.~\ref{fig:theta2}, then $(\phi_i+\phi_j)/2$ can not be a correct definition of $\theta_{ij}$. Instead $(\phi_i+\phi_j)/2-\pi$ gives an appropriate definition of $\theta_{ij}$. }
  \label{fig:theta}
\end{figure}

The phase field $\theta_{ij}$ is defined on the bond $\overline{ij}$, which connects two nearest neighboring sites $i$ and $j$. Therefore it is natural to use the phase fields $\phi_i$ and $\phi_j$, on the site $i$ and site $j$ respectively, to define $\theta_{ij}=\frac{\phi_i+\phi_j}{2}$. However this definition does not guarantee that whenever a closed path encloses a vortex, $\theta_{ij}$ along that path will pick up a $2\pi$ phase as the vortex is winded once. Therefore this $\theta_{ij}$ can not give the correct vortices configuration. It is incorrect whenever a vortex branch cut is crossed. To see this clearly, we map the phase field $\phi_i$ along a closed path that encloses a vortex onto a unit circle since $\phi_i$ is defined only modulo $2\pi$, as schematically shown in Fig.~\ref{fig:theta}. In this figure, the blue arc segment corresponds to the bond $\overline{ij}$ on the closed path. Therefore an appropriate $\theta_{ij}$ should be equal to some value of the phase field on this segment. When the bond $\overline{ij}$ does not cross any branch cut, $\theta_{ij}=\frac{\phi_i+\phi_j}{2}$ is indeed on the blue segment and can be a good definition of $\theta_{ij}$, as illustrated in Fig.~\ref{fig:theta1}\,; however, if the bond $\overline{ij}$ crosses a branch cut, we see that $\frac{\phi_i+\phi_j}{2}$, indicated by the red arrow in Fig.~\ref{fig:theta2}, is not on the blue segment and can not be an appropriate definition of $\theta_{ij}$. In this latter case, $\theta_{ij}=\frac{\phi_i+\phi_j}{2}-\pi$ instead can be a good definition, since it falls onto the blue arc segment, as indicated by the blue arrow in Fig.~\ref{fig:theta2}.

Based on these two scenarios, a good definition of $\theta_{ij}$ will be
\begin{gather}
e^{i\theta_{ij}} = e^{i\frac{\phi_i+\phi_j}{2}} \; \mathrm{sgn}[\cos\frac{\phi_i-\phi_j}{2}]
\end{gather}
This definition of $\theta_{ij}$ guarantees that whenever the phase field $\phi_i$ along a closed path crosses a branch cut once, the defined $\theta_{ij}$ crosses the same branch cut once as well. When there are multiple vortices enclosed, we only need to linearly superpose the contributions from each vortex together to the field $\phi_i$ and $\theta_{ij}$ respectively. It is not difficult to see that the above definition of $\theta_{ij}$ is still good in these cases. For our numerical calculation convenience, we rewrite the above definition of $\theta_{ij}$ in a slightly different way
\begin{align}
e^{i\theta_{ij}} &  = \frac{e^{i\phi_i}+e^{i\phi_j}}{|e^{i\phi_i}+e^{i\phi_j}|}
\end{align}

\section{Recursive Green's function method}\label{sec:rGF}
\begin{figure}
  \centering
  \includegraphics[width=0.48\textwidth]{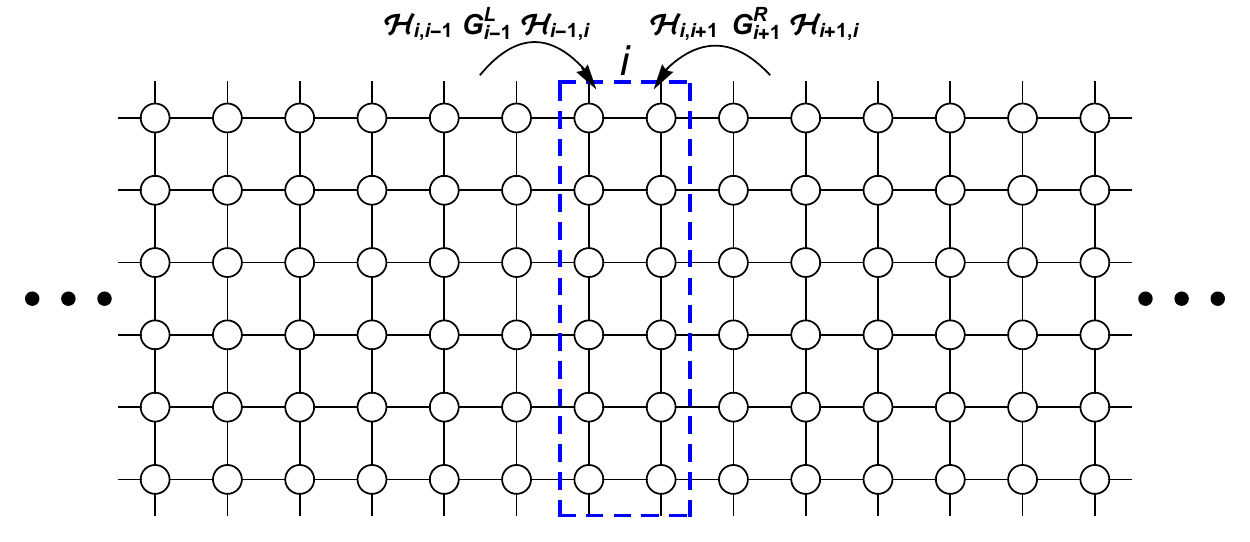}
  \caption{Schematic diagram of the recursive Green's function calculation. The sites enclosed by the blue dashed rectangle define the $i$th principal layer. The exact Green's function $G_i$ has two self-energy contributions from both the left semi-infinite stripe and and the right one. The left stripe is characterized by its surface Green's function $G^{L}_{i-1}$ with the $(i-1)$th layer its surface; while the right one is characterized by another surface Green's function $G^{R}_{i+1}$ with the $(i+1)$th layer its surface.}
  \label{fig:rGF1}
\end{figure}

The recursive Green's function method studies a quasi-one-dimensional system, which is a square lattice with a very long axis of length $La$ along the $x-$direction and a shorter axis of width $Ma$ along the $y-$direction in our problem. The system can be built up recursively in the $x-$direction by connecting many one-dimensional stripes together. Each stripe has a direct coupling only to its nearest neighboring ones. This property is essential for the recursion. In our Hamiltonian $\mathcal{H}$ the third nearest neighbor hopping $t^{\prime\prime}$ provides the farthest direct coupling along the $x-$direction. It connects two sites $2a$ apart. Therefore each stripe necessarily contains two columns of the square lattice sites so that the direct coupling exists only between two adjacent stripes. The blue dashed rectangle in Fig.~\ref{fig:rGF1} shows one such stripe. We define each of such stripes as a principal layer, so each layer contains $2M$ sites.

Our goal is to compute the diagonal matrix elements of the exact Green's function $G$ in order to get the DOS. For this purpose we first calculate $G_i\equiv<i|G|i>$ for each layer $i$. The ket $|i>$ represents a state where the Bogoliubov quasiparticles are found in the $i$th principal layer. It has $4M$ components, of which the first $2M$ ones give the electron part wavefunction, while the rest $2M$ ones define the hole part. Therefore $G_i=G_i(j,j^{\prime})$ is a $4M\times 4M$ matrix with $j,j^{\prime}=1,2,...,4M$. For brevity we will suppress the matrix element indices hereafter, if there is no confusion.

\begin{figure}[ht]
  \centering
  \includegraphics[scale=0.45]{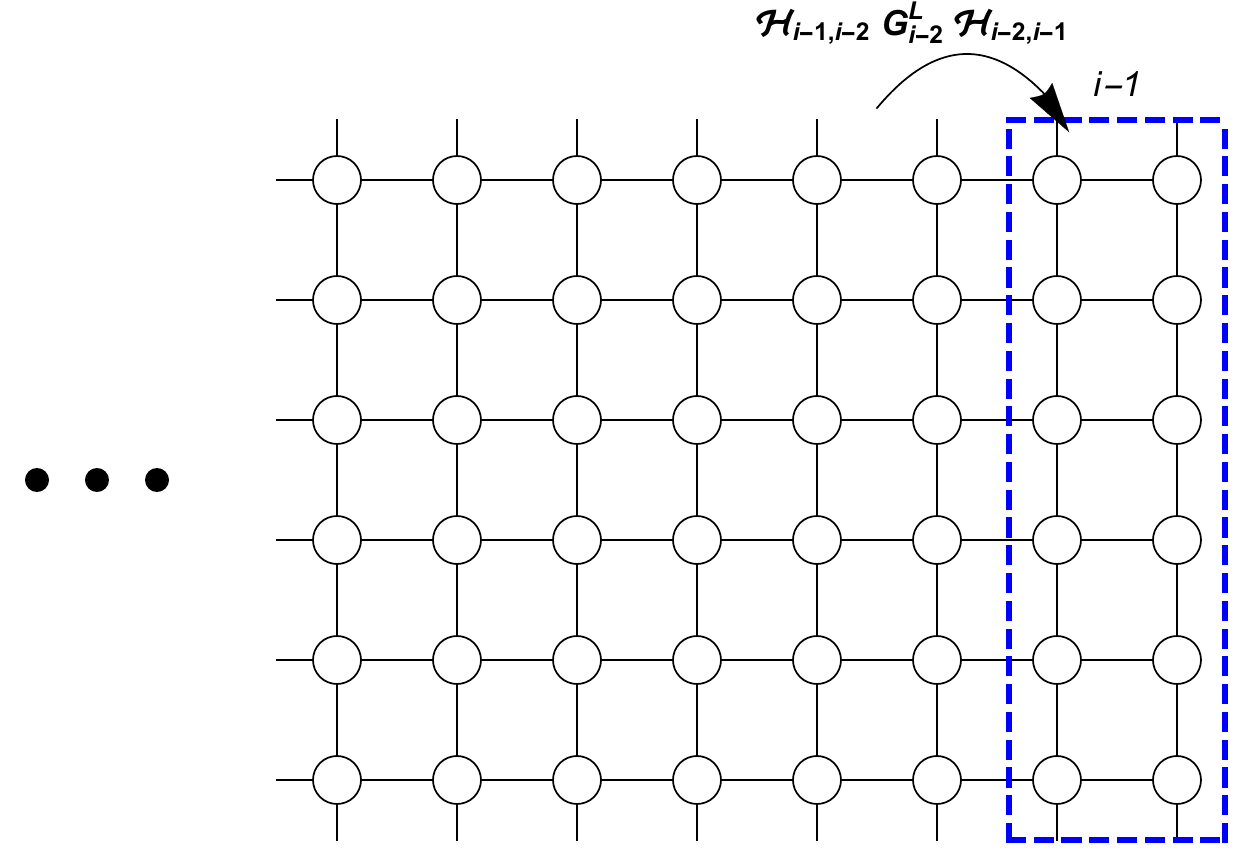}
  \caption{Schematic diagram for the $G^{L}_{i-1}$ computation. The sites enclosed by the blue dashed rectangle belong to the $(i-1)$th principal layer, which is also the surface layer of this semi-infinite stripe.}
  \label{fig:rGF2}
\end{figure}

The exact Green's function $G_i$ can be computed by (for derivations see Ref.\cite{Stauffer1996})
\begin{equation}
 G_i=[{G^{0}_i}^{-1}-\mathcal{H}_{i,i-1} \, G^{L}_{i-1} \, \mathcal{H}_{i-1,i}-\mathcal{H}_{i,i+1} \, G^{R}_{i+1}\, \mathcal{H}_{i+1,i}]^{-1}\,, \label{eq:Gi}
\end{equation}
as schematically shown in Fig.~\ref{fig:rGF1}.
Here $G^{0}_i\equiv[E-<i|\mathcal{H}|i>]^{-1}$ is the bare Green's function of the isolated $i$th principal layer, with the superscript $0$ indicating it is defined as if all other layers are deleted. The matrix $\mathcal{H}_{i,i-1}\equiv <i|\mathcal{H}|i-1>$ contains all the Hamiltonian matrix elements connecting sites in the layer $i-1$ to the layer $i$. Similarly $G^{L}_{i-1}\equiv <i-1|G^{L}|i-1>$ is a matrix defined on the $(i-1)$th principal layer, where $G^{L}$ is the exact Green's function of a subsystem of our original lattice with all layers to the right of the $(i-1)$th layer deleted, as shown in Fig.~\ref{fig:rGF2}. The superscript ``$L$" here means that this subsystem, including a left lead, is extended to the $x=-\infty$. Since the $(i-1)$th layer is the surface of this subsystem, we will call $G^{L}_{i-1}$ the left surface Green's function. Similarly $G^{R}_{i+1}\equiv <i+1|G^{R}|i+1>$ is another surface Green's function of a subsystem of our original lattice with all the layers to the left of the $(i+1)$th layer deleted. Once $G^{L}_{i-1},G^{R}_{i+1}$ are known, $G_i$ can be computed immediately from Eq.~\ref{eq:Gi}.

The central task is then to compute $G^L_{i-1}$ and $G^{R}_{i+1}$. This can be done recursively. Take $G_{i-1}^{L}$ as an example. We start with the leftmost layer $i=1$. There our central system is connected to a semi-infinite lead, which contains infinite number of layers of the same width $M$, numbered by $i=...,-2,-1,0$. We denote this left lead's surface Green's function as $G_{\mathrm{s}}^L$, whose computation will be presented in the following appendix subsection~\ref{subsec:leadsGF}. Then we add the $i=1$st layer of our central system, but not other layers, to this lead so that we get a new semi-infinite stripe. This new stripe has a new surface Green's function denoted as $G^{L}_{1}$, which can be computed from $G_{\mathrm{s}}^{L}$ by
\begin{equation}
  G^{L}_{1} =[{G^{0}_{1}}^{-1}-\mathcal{H}_{1,0} \; G_{\mathrm{s}}^{L} \; \mathcal{H}_{0,1}]^{-1}
\end{equation}
where $\mathcal{H}_{1,0}$ connects sites in the surface layer $i=0$ of the left lead to the $i=1$st layer of our central system. Similarly we can repeat this process by adding one more layer of our central system to the semi-infinite stripe each time, and build up the whole system. In general at an intermediate stage, we may have a semi-infinite stripe, whose surface is, say, the $(i-2)$th layer with a surface Green's function $G^{L}_{i-2}$. Then the $(i-1)$th layer is connected to that stripe to form a new semi-infinite system, which has a new surface Green's function $G^{L}_{i-1}$. And $G^{L}_{i-1}$ can be calculated from $G^{L}_{i-2}$ by the following recursive relation

\begin{equation}
G^{L}_{i-1}=[{G^{0}_{i-1}}^{-1}-\mathcal{H}_{i-1,i-2} \, G^{L}_{i-2} \,\mathcal{H}_{i-2,i-1}]^{-1} \label{eq:L-recursive}
\end{equation}
This is schematically illustrated in Fig.~\ref{fig:rGF2}.

Similarly the right surface Green's function $G^{R}_{i+1}$ can be computed from $G^{R}_{i+2}$ via
\begin{equation}
G^{R}_{i+1} =[{G^{0}_{i+1}}^{-1}-\mathcal{H}_{i+1,i+2} \, G^{R}_{i+2} \, \mathcal{H}_{i+2,i+1}]^{-1}  \label{eq:R-recursive}
\end{equation}
This recursive relation starts with $G^{R}_{L/2}$ at the rightmost layer $i=L/2$ of our central system, where it is connected to another semi-infinite stripe lead extended to $x=\infty$. Note that the central system has only $L/2$ principal layers because each layer contains two columns of the sites, and there are only $L$ columns in total. The layers in this right lead are numbered by $i=L/2+1,L/2+2,....$ We denote the right lead's surface Green's function as $G_{\mathrm{s}}^R$. Then $ G^{R}_{L/2}$ can be computed from $G_{\mathrm{s}}^{R}$ by
\begin{equation}
  G^{R}_{L/2} =[{G^{0}_{L/2}}^{-1}-\mathcal{H}_{L/2,L/2+1} \; G_{\mathrm{s}}^{R} \; \mathcal{H}_{L/2+1,L/2}]^{-1}
\end{equation}
where $\mathcal{H}_{L/2,L/2+1}$ connects our central system to the right lead and contains $t,t^{\prime},t^{\prime\prime}$ only.

\subsection{Surface Green's function $G_{\mathrm{s}}^L,G_{\mathrm{s}}^R$ of the leads}\label{subsec:leadsGF}
$G_{\mathrm{s}}^L,G_{\mathrm{s}}^R$ can be computed by solving a self-consistent $2\times 2$ matrix equation. We now give a detail discussion on how to compute $G_{\mathrm{s}}^L$, but only briefly mention the final results for $G_{\mathrm{s}}^R$ at the end.

The left lead Hamiltonian contains only the hopping parameters $t,t^{\prime},t^{\prime\prime}$

\begin{align}
H_{\mathrm{lead}}& =\sum_{i=-\infty}^{0} \sum_{j=1}^M \{ -t[ c_{i-1,j}^{\dagger}c_{i,j}+ c_{i,j+1}^{\dagger}c_{i,j}] \nonumber \\
& +t^{\prime}[ c_{i-1,j+1}^{\dagger}c_{i,j}+ c_{i-1,j-1}^{\dagger}c_{i,j}] \nonumber \\
&  -t^{\prime\prime}[ c_{i-2,j}^{\dagger}c_{i,j}+ c_{i,j+2}^{\dagger}c_{i,j}]+\mathrm{h.c.}  -\mu \, c_{i,j}^{\dagger}c_{i,j} \}
\end{align}

To be compatible with our central system Hamiltonian, which contains superconductivity, our lead Hamiltonian should have both an electron part and a hole part so that the full Hamiltonian $\mathcal{H}_{\mathrm{lead}}$ is
\begin{gather}
\mathcal{H}_{\mathrm{lead}}=\left( \begin{array}{cc} H_{\mathrm{lead}} & 0 \\ 0 & -H_{\mathrm{lead}} \end{array} \right) \,.
\end{gather}
Correspondingly the surface Green's function takes a block diagonal form
\begin{align}
& G_{\mathrm{s}}(E^{+})   \left( \begin{array}{cc} [E^{+}-H_{\mathrm{lead}}]^{-1} & 0 \\ 0 & [E^{+}+H_{\mathrm{lead}}]^{-1} \end{array} \right)
\end{align}
where for brevity we have introduced $E^{+}=E+i\delta$. We will denote the two diagonal terms as $G_{ee}=[E^{+}-H_{\mathrm{lead}}]^{-1}$ and $G_{hh}=[E^{+}+H_{\mathrm{lead}}]^{-1}$. Apparently $G_{\mathrm{hh}}$ can be obtained from $G_{ee}$ by simple substitutions: $\{t,t^{\prime},t^{\prime\prime},\mu\} \Rightarrow \{-t,-t^{\prime},-t^{\prime\prime},-\mu\}$. Therefore we only need to discuss how to compute $G_{ee}$.

Because of the periodic boundary condition along the $y-$direction, we can decompose $G_{ee}( E^{+})$ into different momentum $k_y$ channels
\begin{align}
G_{ee}(E^{+})=\sum_{k_y} |\chi_{k_y}><\chi_{k_y}| \; g(k_y,E^{+}) \label{eq:Geegky}
\end{align}
with $ |\chi_{k_y}>=\sum_{j=1}^{M}\frac{e^{i k_y ja}}{\sqrt{M}} \, |j> \, ,$
and $k_y=\frac{2n\pi}{Ma}$ with $n=1,2,3,...,M$. Each channel is described by a semi-infinite one dimensional chain effective Hamiltonian $H_{\mathrm{eff}}(k_y) \equiv <\chi_{k_y}|H_{\mathrm{lead}}|\chi_{k_y}>$, given by
\begin{align}
  & H_{\mathrm{eff}}(k_y)  =\sum_{i=-\infty}^{0} \{ (- 2\, t\, \cos k_y-2\,t^{\prime\prime}\, \cos 2k_y-\mu) \, c^{\dagger}_i c_i  \nonumber \\
 & +[(-t+2\, t^{\prime} \cos k_y) \, c^{\dagger}_{i}c_{i-1} -t^{\prime\prime} c^{\dagger}_{i}c_{i-2} + \mathrm{h.c.}] \}.
\end{align}
And $g(k_y,E^{+})$ is the corresponding surface Green's function of this one dimensional chain. 

To compute $g(k_y,E^{+})$ we group every two adjacent cites $( c_{i-1}^{\dagger} \; ,   c_{i}^\dagger)$ of the one dimensional chain together into a cell, indexed by the cell number $n$, so that $H_{\mathrm{eff}}(k_y)$ can be rewritten in a form such that direct couplings exist only between two nearest neighboring cells
\begin{widetext}
\begin{align}
H_{\mathrm{eff}}(k_y)= & \sum_{n=-\infty}^{0} (\begin{array}{ll} c_{2n-1}^{\dagger} \; , &  c_{2n}^\dagger  \end{array}) \left[\begin{array}{cc} -t^{\prime\prime} & -t+2 t^{\prime}\cos k_y \\ 0 & -t^{\prime\prime} \end{array}\right]\left(\begin{array}{l} c_{2n-3} \\ c_{2n-2} \end{array}\right) + \mathrm{h.c.} \nonumber \\
+ & \sum_{n=-\infty}^{0} (\begin{array}{ll} c_{2n-1}^{\dagger} \; , &  c_{2n}^\dagger  \end{array}) \left[\begin{array}{cc} -2 t \cos k_y-2 t^{\prime\prime}\cos 2k_y-\mu & -t+2t^{\prime}\cos k_y \\ -t+2 t^{\prime}\cos k_y & -2 t\cos k_y -2 t^{\prime\prime}\cos 2k_y -\mu \end{array}\right]\left(\begin{array}{l} c_{2n-1} \\ c_{2n} \end{array}\right)
 \label{eq:Heff}
\end{align}
\end{widetext}
Since $g(k_y,E^{+})$ is a surface Green's function, it should satisfy the same recursive relation given in Eq.~\eqref{eq:L-recursive}, which is rewritten here as
\begin{align}
 g=[{G^{0}_{0}}^{-1}-[H_{\mathrm{eff}}]_{0,-1} \, G_{-1}^{L} \, [H_{\mathrm{eff}}]_{-1,0}]^{-1} \label{eq:surf-gR}
\end{align}
The only difference from there is now all the matrix elements are defined between different cells instead of layers. For clarity we have suppressed the $k_y$ and $ E^{+}$ dependence of all the quantities in this equation. $G_{0}^0$ is the bare Green's function of the isolated single cell $n=0$. Because each cell contains two sites, $G^{0}_{0}$ is a $2\times 2$matrix, given by
\begin{widetext}
\begin{align}
     & { G^{0}_{0}}^{-1}  \equiv E^{+}-[H_{\mathrm{eff}}]_{0,0}=E^{+}-   \left[\begin{array}{cc} -2 t \cos k_y-2 t^{\prime\prime}\cos 2k_y-\mu & -t+2t^{\prime}\cos k_y \\ -t+2 t^{\prime}\cos k_y & -2 t\cos k_y -2 t^{\prime\prime}\cos 2k_y -\mu \end{array}\right]\,.
\end{align}
\end{widetext}

Similarly the effective hopping matrices between the cell $n=0$ and cell $n=-1$ can be read off directly from Eq.~\eqref{eq:Heff}
\begin{align}
      [H_{\mathrm{eff}}]_{0,-1} & = \left[\begin{array}{cc} -t^{\prime\prime} & -t+2 t^{\prime}\cos k_y \\ 0 & -t^{\prime\prime} \end{array}\right], \\ [H_{\mathrm{eff}}]_{-1,0} & =[H_{\mathrm{eff}}]_{0,-1}^{\dagger}
\end{align}
By definition $G_{-1}^{L}$ in Eq.~\eqref{eq:surf-gR} is the surface Green's function of the same chain but with the cell $n=0$ deleted. However, since the chain is semi-infinite, deleting the surface cell only gives another identical semi-infinite chain. Therefore $G_{-1}^{L}$ should be the same as $g$. Then Eq.~\eqref{eq:surf-gR} becomes a self-consistent equation of $g$ as
\begin{widetext}
\begin{align}
g^{-1} &  =\left[\begin{array}{cc} E^{+} +2 t\cos k_y + 2 t^{\prime\prime}\cos 2k_y+\mu & t-2t^\prime \cos k_y \\ t- 2t^\prime \cos k_y & E^{+} +2 t \cos k_y + 2 t^{\prime\prime}\cos 2k_y +\mu\end{array} \right] \nonumber \\
 & - \left[\begin{array}{cc} -t^{\prime\prime} &  -t+2 t^{\prime}\cos k_y \\ 0  & -t^{\prime\prime} \end{array}\right] g \left[\begin{array}{cc} -t^{\prime\prime} & 0 \\ -t+2 t^{\prime}\cos k_y  & -t^{\prime\prime} \end{array}\right].
\end{align}
\end{widetext}
With this $2\times 2$ matrix equation, for each $k_y$, we solve for $g$ numerically by iterations until the results converge. Then the computed $g(k_y,E)$ is substituted back into Eq.~\eqref{eq:Geegky} of $G_{\mathrm{ee}}(E^{+})$ to get $G_{\mathrm{s}}^{L}$.

Similar derivations can be carried out for the right lead Green's function $G^{R}_{\mathrm{s}}$. It turns out $G_{\mathrm{s}}^{R}=(G_{\mathrm{s}}^L)^{\mathrm{T}}$, where $\mathrm{T}$ is the transpose operation. This result is a manifestation of the fact that the two semi-infinite leads can be connected to each other by a reflection symmetry operation along the $x-$direction.

\section{The ansatz $(\frac{\xi}{r_{\mathrm{eff}}})^{q}=\sum\limits_{n} (\frac{\xi}{r_{n}})^{q}$}\label{sec:ansatz}
The pairing amplitude on the bond, that connects two nearest neighboring sites $\vec{r}_i$ and $\vec{r}_j$, is calculated by the following ansatz:
\begin{gather}
|\Delta_{ij}|=\Delta \frac{r_{\mathrm{eff}}}{\sqrt{\xi^2+r_{\mathrm{eff}}^2}} \label{eq:deltaij}
\end{gather}
with $r_{\mathrm{eff}}$ given by
\begin{gather}
(\frac{\xi}{r_{\mathrm{eff}}})^{q}=\sum_{n=1}^{N_v} (\frac{\xi}{r_{n}})^{q} \label{eq:reff}
\end{gather}
where $r_n=|\frac{\vec{r}_i+\vec{r}_j}{2}-\vec{R}_n|$ is the distance from the bond center $\frac{\vec{r}_i+\vec{r}_j}{2}$ to the $n$th vortex center $\vec{R}_n$, $q$ is some positive number, and $N_v$ is the total number of vortices.

If we consider a special case that there is only one vortex, for instance the $n$th vortex, then Eq.~\eqref{eq:reff} is reduced to $r_{\mathrm{eff}}=r_n$, and
\begin{gather}
 |\Delta_{ij}| =\Delta \frac{r_n}{\sqrt{r_n^2+\xi^2}}
\end{gather}
In other words, we can define the pairing amplitude $|\Delta_n|$ for the case when only the $n$th vortex is present as follows
\begin{align}
|\Delta_n|\equiv \Delta \frac{r_n}{\sqrt{r_n^2+\xi^2}} \label{eq:deltan}
\end{align}
so that $|\Delta_{ij}|=|\Delta_n|$.

When more than the $n$th vortex is present, $|\Delta_{ij}|$ should become smaller than $|\Delta_n|$. This requires $r_{\mathrm{eff}}<r_n$ because $|\Delta_{ij}|$ is an increasing function of $r_{\mathrm{eff}}$, as seen in Eq.~\eqref{eq:deltaij}. We sum the contributions from each vortex to $|\Delta_{ij}|$ simply by adding the $q$th inverse moment($q>0$) of all $r_n$ together to define an effective distance $r_{\mathrm{eff}}$ as in Eq.~\eqref{eq:reff}. Using the $q$th inverse moment, instead of the $q$th moment guarantees that $r_{\mathrm{eff}}<r_n$ when there is more than one vortex. Furthermore it ensures that the terms $(\frac{\xi}{r_{n}})^{q}$ with small $r_n$ on the right hand side of Eq.~\eqref{eq:reff} contribute more significantly than those with larger $r_n$. This is consistent with the physical intuition that vortices nearby are more important in determining $r_{\mathrm{eff}}$, and therefore $|\Delta_{ij}|$, than those that are far away. Also when there are more vortices present, $N_v$ becomes larger and the resultant $|\Delta_{ij}|$ from Eq.~\eqref{eq:reff} becomes smaller. This again agrees with our expectation.

The $|\Delta_{ij}|$ defined above increases monotonically with the parameter $q$ for a given vortices configuration. To see this we only need to show $r_{\mathrm{eff}}$ increases with $q$. For that purpose we can rewrite Eq.~\eqref{eq:reff} as follows
\begin{align}
\log\frac{r_{\mathrm{min}}}{r_{\mathrm{eff}}} &= \frac{1}{q}\log\{1+\sideset{}{'}\sum_{n} (\frac{r_{\mathrm{min}}}{r_{n}})^{q}\} \label{eq:reff2}
\end{align}
where $r_{\mathrm{min}}=\mathrm{min}\{r_n\}$ is the distance between the closest vortex and the bond, and the prime sign in the summation means this closest vortex is excluded. The right hand side of Eq.~\eqref{eq:reff2} is a monotonic decreasing function of $q$ because in the summation each $\frac{r_{\mathrm{min}}}{r_{n}}<1$. Therefore $r_{\mathrm{eff}}$ increases monotonically with $q$, so does $|\Delta_{ij}|$.
\begin{figure}[htp]
 \centering
 \includegraphics[ width=0.4\textwidth]{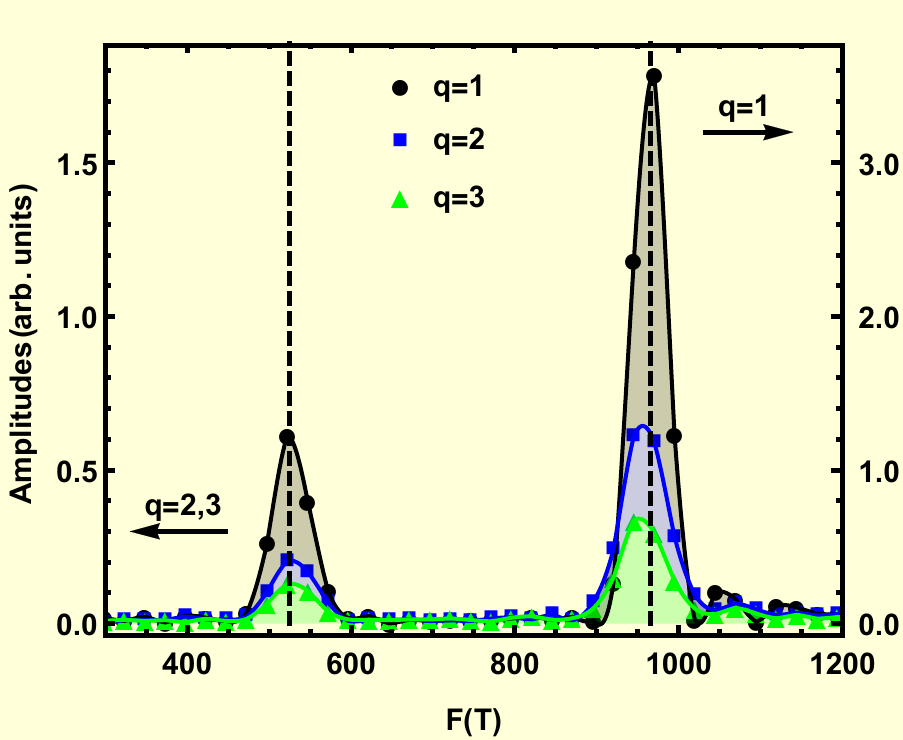}
 \caption{Oscillation spectrum of the DOS for the two fold DDW order model with different $q$ values but a fixed $\Delta$. Note that the vertical axis scale for $q=1$ is different from those for other $q$ values. The two vertical dashed lines mark the two frequencies $F_e=525 \mathrm{T},F_h=966\mathrm{T}$. }
 \label{fig:dos-tpp-2fddw-q12345-d10}
\end{figure}

An appropriate value of $q$ can not be determined without solving the whole problem self-consistently, therefore we performed simulations for different $q$ to see if our conclusions depend on $q$ or not. One example data of the oscillation spectrum for the two-fold DDW order case is shown in Fig.~\ref{fig:dos-tpp-2fddw-q12345-d10}. From this figure, we observe that the oscillation amplitude decreases as $q$ is increased from $q=1$ to $q=3$. This is consistent with the analyses that $|\Delta_{ij}|$ is a monotonic increasing function of $q$, since larger $q$ gives larger $|\Delta_{ij}|$, which means stronger vortex scattering and therefore stronger suppression of the oscillation amplitudes.

Although the oscillation amplitudes can depend significantly on $q$, the oscillation frequencies remain unaffected by varying the $q$ values, therefore the conclusion of Onsager's relation being robust against the vortex scattering does not depend on the value of $q$.

\section{Check of the results in Ref.\cite{Chen2009}}\label{sec:patrick}
We have checked the Fig.1 and Fig.3 of Ref.\cite{Chen2009}, using the same parameter sets, and find our conclusions remain the same. For both these two cases, the normal state, without magnetic field, can be described by the following Hamiltonian
\begin{align}
H & =-t \sum_{<i,j>} c_i^{\dagger}c_j \, +t^{\prime} \sum_{<<i,j>>} c_i^{\dagger} c_j  \nonumber \\
& +\sum_{<i,j>}i\; (-1)^{x_i+y_i} \eta_{ij}\frac{W_0}{4} c_i^{\dagger}c_j -\mu \sum_{i} c_i^{\dagger} c_i \, . \label{eq:PALeeH}
\end{align}
In this Hamiltonian the third term is a two-fold DDW order(or the staggered flux state order), and $\eta_{ij}=\pm1$ is again the local $d-$wave symmetry factor.
\begin{enumerate}
\item First consider the Fig.1 of Ref.\cite{Chen2009}. We use the same parameters $t=1,t^{\prime}=0.3 t, W_0=1.0 t, \mu=-0.949 t$. The normal state Fermi surface consists of four hole pockets with an area $\frac{A_F}{(2\pi/a)^2}\approx 2.5\%$ each. Fig.~\ref{fig:PALeeFig1-q=1} shows the computed DOS. We see there is no noticeable shift in the oscillation frequency when the vortex scattering is present.
  \begin{figure}[htp]
  \centering
  \includegraphics[width=0.4\textwidth]{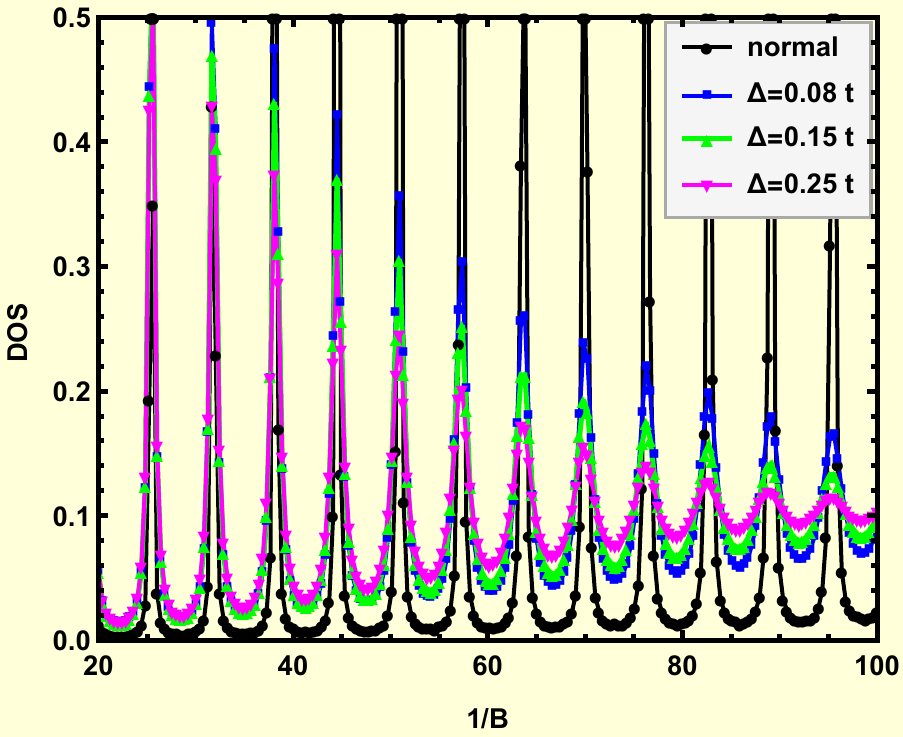}
  \caption{DOS oscillation for $t=1,t^{\prime}=0.3t,W_0=1.0 t,\mu=-0.949 t$. The unit for the field $B$ is $\frac{\Phi_0}{2\pi a^2}$, with $\Phi_0=hc/e$ the full flux quantum and $a$ the lattice spacing. And the DOS unit is $\mathrm{states}/t$ . In the legends the ``normal"  means $\Delta=0$. The lattice size is $L=2000,M=80$, and in the Eq.~\eqref{eq:reff} of $r_\mathrm{eff}$, $q=1$ rather than $q=2$ has been chosen here. }
  \label{fig:PALeeFig1-q=1}
\end{figure}


\item Then consider the Fig.3 of Ref.\cite{Chen2009}. In this case, the normal state does not have DDW, so $W_0=0$. For $ t=1,t^{\prime}=0.14 t, \mu=-2.267 t$, the obtained Fermi surface contains only a large hole pocket with an area $\frac{A_F}{(2\pi/a)^2}\approx 14\%$ at the Brillouin zone center. Fig.~\ref{fig:PALeeFig3-q=2} shows the corresponding DOS results. We see the oscillation amplitude gets heavily damped as $\Delta$ increases. Moreover, a small frequency shift $\delta F/F \approx 2\%$ becomes noticeable. However, this is different from a large $30\%$ shift found in Ref.\cite{Chen2009}. Also this $2\%$ shift does not contradict our previous conclusion of no noticeable frequency shift. Because the shift here is obtained at magnetic fields that are larger than the experimentally applied fields($\sim 50\mathrm{T}$) by an order of magnitude. In Fig.~\ref{fig:PALeeFig3-q=2}, $\frac{1}{B}=10$ corresponds to $ B=\frac{1}{10}\, \frac{\Phi_0}{2\pi a^2}\approx 450\mathrm{T}$, since $\Phi_0/2\pi a^2\approx 4500\mathrm{T}$ if we take $a=3.83\AA$ for $\mathrm{YBCO}$.
    
    \begin{figure}[htp]
    \centering
     \includegraphics[width=0.4\textwidth]{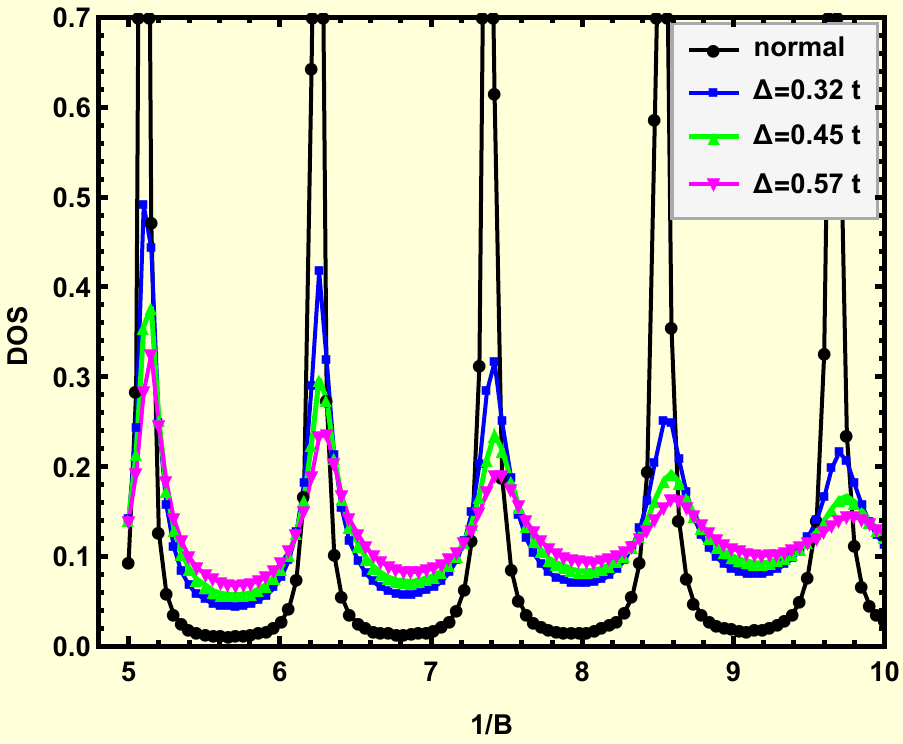}
     \caption{ DOS oscillation for $t=1,t^{\prime}=0.14t,W_0=0,\mu=-2.267 t$. In this simulation the system size is $L=1000,M=100$. }
     \label{fig:PALeeFig3-q=2}
 \end{figure}
 
\end{enumerate}
%


\end{document}